\documentclass[fleqn,usenatbib]{mnras}

\usepackage{newtxtext,newtxmath}
\usepackage[T1]{fontenc}

\DeclareRobustCommand{\VAN}[3]{#2}
\let\VANthebibliography\thebibliography
\def\thebibliography{\DeclareRobustCommand{\VAN}[3]{##3}\VANthebibliography}

\usepackage{graphicx}	% Including figure files
\usepackage{amsmath}	% Advanced maths commands

\newcommand\HI{\ion{H}{I}\,} % ionized hydrogen
\newcommand\HII{\ion{H}{II}\,} % ionized hydrogen
\newcommand\OIII{[\ion{O}{III}]\,} 
\newcommand\OIIIl{[\ion{O}{III}]$\lambda$\,} 
\newcommand\OII{[\ion{O}{II}]\,} % ionized hydrogen
\newcommand{\CIII}{{C\,{\sc iii}]}}

\newcommand{\CIV}{{C\,{\sc iv}}}

\newcommand{\HeI}{{He\,{\sc i}\,}}
\newcommand{\HeII}{{He\,{\sc ii}\,}}
\newcommand{\HeIIl}{{He\,{\sc ii}\,$\lambda$}}
\newcommand{\NII}{{[N\,{\sc ii}]\,}}
\title[Nature of GS9422]{Resolving the nature and putative nebular emission of GS9422: \\an obscured AGN without exotic stars}

\author[Tacchella et al.]{Sandro Tacchella,$^{1,2}$\thanks{E-mail: st578@cam.ac.uk}
William McClymont,$^{1,2}$
Jan Scholtz,$^{1,2}$
Roberto Maiolino,$^{1,2}$
Xihan Ji,$^{1,2}$
\newauthor
Natalia C. Villanueva,$^{1,2}$
St\'ephane Charlot,$^{3}$
Francesco D'Eugenio,$^{1,2}$
Jakob M. Helton,$^{4}$
\newauthor
Christina C. Williams,$^{5}$
Joris Witstok,$^{1,2}$
Rachana Bhatawdekar,$^{6}$
Stefano Carniani,$^{7}$
Jacopo Chevallard,$^{8}$
\newauthor
Mirko Curti,$^{9}$
Kevin Hainline,$^{4}$
Zhiyuan Ji,$^{4}$
Benjamin D.\ Johnson,$^{10}$
Joel Leja,$^{11,12,13}$
Yijia Li,$^{11,13}$
\newauthor
Michael V. Maseda,$^{14}$
D\'avid Pusk\'as,$^{1,2}$
Marcia Rieke,$^{4}$
Brant Robertson,$^{15,16}$
Irene Shivaei,$^{17}$
\newauthor
Maddie S. Silcock,$^{18}$
Charlotte Simmonds,$^{1,2}$
Hannah \"Ubler,$^{1,2}$
Christopher N. A. Willmer,$^{4}$
Chris Willott$^{19}$
\\
\\
\emph{\normalsize Affiliations are listed at the end of the paper}
}

\begin{document}
\label{firstpage}
\pagerange{\pageref{firstpage}--\pageref{lastpage}}
\maketitle

\begin{abstract}
Understanding the sources that power nebular emission in high-redshift galaxies is fundamentally important not only for shedding light onto the drivers of reionisation, but to constrain stellar populations and the growth of black holes. Here we focus on an individual object, GS9422, a galaxy at $z_{\rm spec}=5.943$ with exquisite data from the JADES, JEMS and FRESCO surveys, including 14-band JWST/NIRCam photometry and deep NIRSpec prism and grating spectroscopy. We map the continuum emission and nebular emission lines across the galaxy on 0.2-kpc scales. GS9422 has been claimed to have nebular-dominated continuum and an extreme stellar population with top-heavy initial mass function. We find clear evidence for different morphologies in the emission lines, the rest-UV and rest-optical continuum emission, demonstrating that the full continuum cannot be dominated by nebular emission. While multiple models reproduce the spectrum reasonably well, our preferred model with a type-2 active galactic nucleus (AGN) and local damped Ly-$\alpha$ (DLA) clouds can explain both the spectrum and the wavelength-dependent morphology. The AGN powers the off-planar nebular emission, giving rise to the Balmer jump and the emission lines, including Ly-$\alpha$, which therefore does not suffer DLA absorption. A central, young stellar component dominates the rest-UV emission and -- together with the DLA clouds -- leads to a spectral turn-over. A disc-like, older stellar component explains the flattened morphology in the rest-optical continuum. We conclude that GS9422 is consistent with being a normal galaxy with an obscured, type-2 AGN -- a simple scenario, without the need for exotic stellar populations.
\end{abstract}


%%%%%%%%%%%%%%%%%%%%%%%%%%%%%%%%%%%%%%%%%%%%%%%%%%

%%%%%%%%%%%%%%%%% BODY OF PAPER %%%%%%%%%%%%%%%%%%

\section{Introduction}
\label{sec:introduction}

Nebular emission both in the form of lines and its continuum provides crucial information on the fundamental physical processes occurring within the galaxy \citep[e.g.,][]{kewley19_review}. The emission lines can be used as diagnostics for the abundance of elements in the gas, the gas density, and the pressure of the interstellar medium \citep[ISM; e.g.,][]{kewley19, maiolino19}. Different sources and mechanisms can power the nebular emission, including shocks (from, e.g., winds driven by stars or black holes, or galaxy-galaxy interactions) or radiation from stars and accreting supermassive black holes. The ionising spectrum of stars depends on a wide range of parameters, including the stellar age, metallicity, the number of young stars (star-formation rate [SFR]), the stellar binarity fraction, the rate of rotation in stars, and the initial mass function \citep[IMF; e.g.,][]{leitherer99, bruzual03, choi17, lecroq24}. The emission from an accreting supermassive black hole is complex, depending significantly on viewing-angle effects, details of the dust geometry around the accretion disc, and black hole properties such as mass and spin \citep[e.g.,][]{greene05, reynolds21}. Even if the ionising source is known, radiative transfer effects are important and high-energy photons can -- in addition to ionising gas -- also be absorbed by gas or escape the galaxy \citep[e.g.,][]{steinacker13, katz18, tacchella22_Halpha}. 

JWST has opened up a new window: we are now able to study the rest-frame optical emission of galaxies and their black holes out to redshift $z>10$. JWST/NIRSpec is the first near-IR spectrograph with a high-spectral resolution ($R>1000$) in space. Tracing the intensities and shapes of key optical emission lines allowed us to constrain, in galaxies at $z\approx 4-11$, the black hole population and their growth rates \citep[e.g.,][]{harikane23, kocevski23, kokorev23, larson23, maiolino24_bh, maiolino24_gnz11, ubler23, scholtz23, greene24}, the ISM conditions \citep[e.g.,][]{brinchmann23, cameron23_el}, the early chemical enrichment \citep[e.g.,][]{curti23_ero, curti24_Z, deugenio24_C, tacchella23_metal}, the stellar populations \citep[e.g.,][]{schaerer22, maiolino24_popIII, cameron24, looser23_pop}, the outflow properties \citep[e.g.,][]{belli24, carniani24_outflow, deugenio24_psb, maiolino24_gnz11, davies24}, the ionised gas kinematics \citep[e.g.,][]{de-graaff24_kin, nelson24, ubler24}, and the production and escape of ionising photons \citep[e.g.,][]{endsley24, saxena24, simmonds23_jems, simmonds24, tang23}. While emission lines can dominate the photometry for actively star-forming galaxies and can therefore be easily detected \citep{stark13, faisst19, tacchella22_highz, trussler23, williams23_jems}, the nebular continuum is more challenging to identify. 

The nebular continuum arises from three different processes \citep{osterbrock89, reines10, byler17}: ($i$) two-photon emission, ($ii$) free-bound emission (recombination continuum), and ($iii$) free-free emission (Bremsstrahlung). The two-photon continuum is the result of a bound-bound process, where the excited 2s state of hydrogen decays to the 1s state by the simultaneous emission of two photons. The two-photon continuum can be important in the ultra-violet (UV), peaking at rest-frame wavelength of $\lambda_{\rm rest}\approx1430$ \AA, but is typically smaller in magnitude than the stellar UV emission; additionally, the two-photon continuum has low critical density ($7\times 10^3~\mathrm{cm}^{-3}$; \citealt{mewe86}), hence can only be observed in low-density ionised regions. The free-bound continuum is produced when a free electron recombines into an excited level of hydrogen, followed by the radiative cascade that produces recombination lines. The resultant continuum spectrum has a sharp jump at the ionisation energy, followed by continuous emission to higher energies. For the hydrogen Balmer line series (recombinations to level 2), this gives rise to the Balmer jump at $\lambda_{\rm rest}\approx3645$ \AA. The free–free continuum, which is the result of a free electron scattering off of an ion or proton, produces a roughly power-law distribution of photon energies and is only important at longer wavelengths (beyond 1 $\mu$m), but typically outshone by dust-reprocessed stellar emission. 

Nebular continuum emission is the strongest at high ionisation parameters and low metallicity \citep[e.g.,][]{byler17}. Spectroscopic data of \HII regions or galaxies in the local Universe show the presence of Balmer jumps \citep{peimbert69, guseva06, hagele06}. \citet{reines10} found that the nebular continuum contributed significantly ($\sim40\%$) to the NUV-NIR broadband flux of young ($<3$ Myr) massive star clusters. Importantly, the harder ionising spectra associated with young stellar populations, shocks or an AGN further enhance the nebular continuum strength. It is therefore not surprising that nebular continuum emission (incl. Balmer and Paschen jumps) have been detected in the photometry and spectroscopy of local AGN \citep{baldwin75, grandi82, kovacevic14, guo22_paschenjump}. Furthermore, AGN monitoring results have suggested a substantial or even dominant contribution from nebular continuum emission to AGN optical variability \citep{chelouche19, cackett22, netzer22}.

At higher redshifts, finding and constraining nebular continuum emission is more challenging. JWST photometry, in combination with spectral energy distribution (SED) modelling, indicates the presence of nebular continuum emission  \citep{endsley24, tacchella23_metal, topping24_slope}. To fully pin down the nebular continuum, deep spectroscopic data is necessary. Recently, \citet{cameron24} reported the first clear spectroscopic detection of nebular continuum in a galaxy at $z>4$ (JADES-GS+53.12175-27.79763; hereafter GS9422) and interpret this as an indication of a top-heavy IMF. They claim that the rest-frame UV-to-optical spectrum, which shows a clear Balmer jump and a steep turnover in the UV continuum, is dominated by nebular continuum emission. By interpreting the UV turnover as caused by two-photon emission, they derive an ionising emissivity that is $\gtrsim10\times$ than that of a typical star-forming galaxy at this epoch. The weak HeII emission makes them disfavour an origin from an AGN or X-ray binaries, concluding that this galaxy contains star clusters dominated by low-metallicity stars of $\gtrsim50~M_{\odot}$, where the IMF is $10-30\times$ more top-heavy than typically assumed. In contrast, \citet{scholtz23} selected GS9422 to be an AGN based on line ratios involving HeII$\lambda1640$ and HeII$\lambda4686$. 

In this work, we use detailed imaging and spectroscopic data for this galaxy, GS9422, to provide further insights into the nature of the putative nebular emission first discussed by \citet{cameron24}. The inclusion in our analysis of 14-band imaging data (including medium bands), allows us to confirm the presence of a Balmer jump, while also finding clear evidence for different morphological features in the nebular emission lines, rest-optical and rest-UV continuum emissions. The morphological information challenges the idea that the whole NIRSpec spectrum is dominated by nebular emission because, in that case, one would expect continuum and emission lines to be co-spatial. Furthermore, key emission-line diagnostics indicate that this galaxy lies in a parameter space that is common for AGN. While we explore a range of different models (including pure stellar emission, pure AGN emission, and a combination of both), our preferred model includes a galaxy with a type-2 AGN and damped Ly-$\alpha$ (DLA) clouds. This model can explain both the spectrum and the wavelength dependence of the morphology. The AGN powers the off-planar nebular emission, giving rise to the Balmer jump and emission lines. A centrally concentrated, young stellar component dominates the rest-UV emission, and together with the DLA clouds leads to a turn-over of the spectrum towards shorter wavelengths. The older stellar component in a disc-like configuration can explain the observed flattened morphology in the rest-optical. In summary, we find that GS9422 is consistent with being a normal galaxy with an AGN -- a simple scenario that can explain its properties, without the need for stellar populations with extreme IMFs.

This paper is structured as follows. Section~\ref{sec:observation} summarises the NIRCam and NIRSpec observations and presents the basic properties of GS9422. Section~\ref{sec:morphology} analyses the morphology of GS9422. In Section~\ref{subsec:el_diag}, we discuss UV and optical emission line diagnostics diagrams. Section~\ref{sec:sed} focuses on the SED modelling and discusses this in the context of the morphology analysis. We summarise and conclude in Section~\ref{sec:conclusions}.

Throughout this work, we use the AB magnitude system and assume the Planck18 flat $\Lambda$CDM cosmology \citep{planck-collaboration20} with $\Omega_m=0.315$ and $H_0=67.4$ km/s/Mpc. In this cosmology $1''$ corresponds to a transverse distance of 5.863 proper kpc at spectroscopic redshift of GS9422 of $z=5.943$.

\section{GS9422: observations \& basic properties}
\label{sec:observation}

\begin{figure}
	\includegraphics[width=\linewidth]{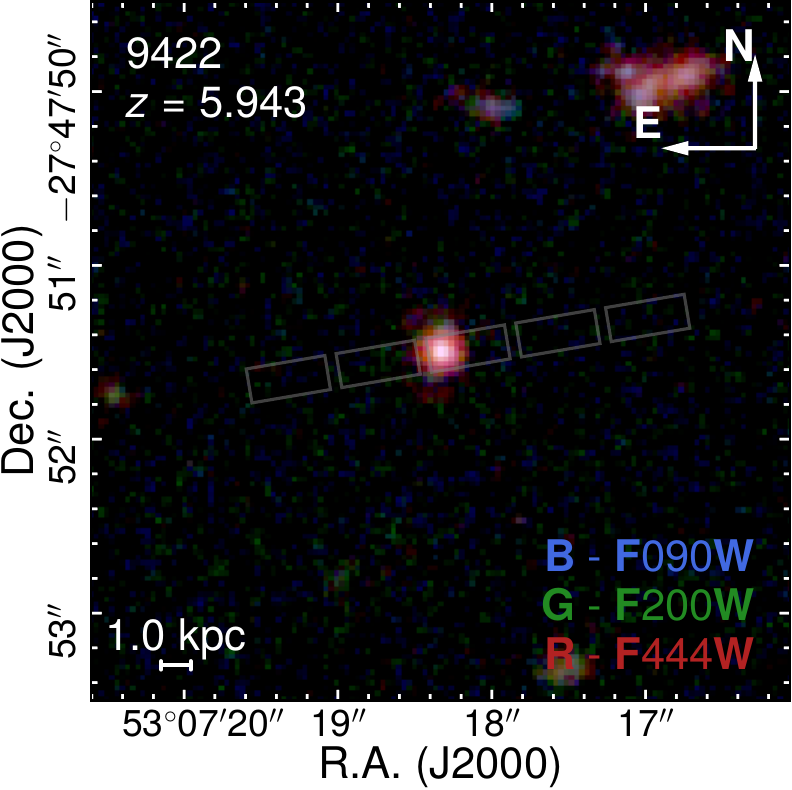}
    \caption{RGB colour composite of GS9422, highlighting the position of the three NIRSpec/MSA shutters ($\mathrm{RA}=53.121757^{\circ}$ and $\mathrm{Dec}=-27.797638^{\circ}$; the five-shutter overlay covers the nodding sequence). The colours correspond to the NIRCam filters F090W (B: blue), F200W (G: green), and F444W (R: red). GS9422 lies at a spectroscopic redshift of $z_{\rm spec}=5.943$.}
    \label{fig:slit_position}
\end{figure}

% \begin{figure}
% 	\includegraphics[width=\linewidth]{figures/test.pdf}
%     \caption{Testing larger cutout}
% \end{figure}

\begin{figure*}
	\includegraphics[width=\textwidth]{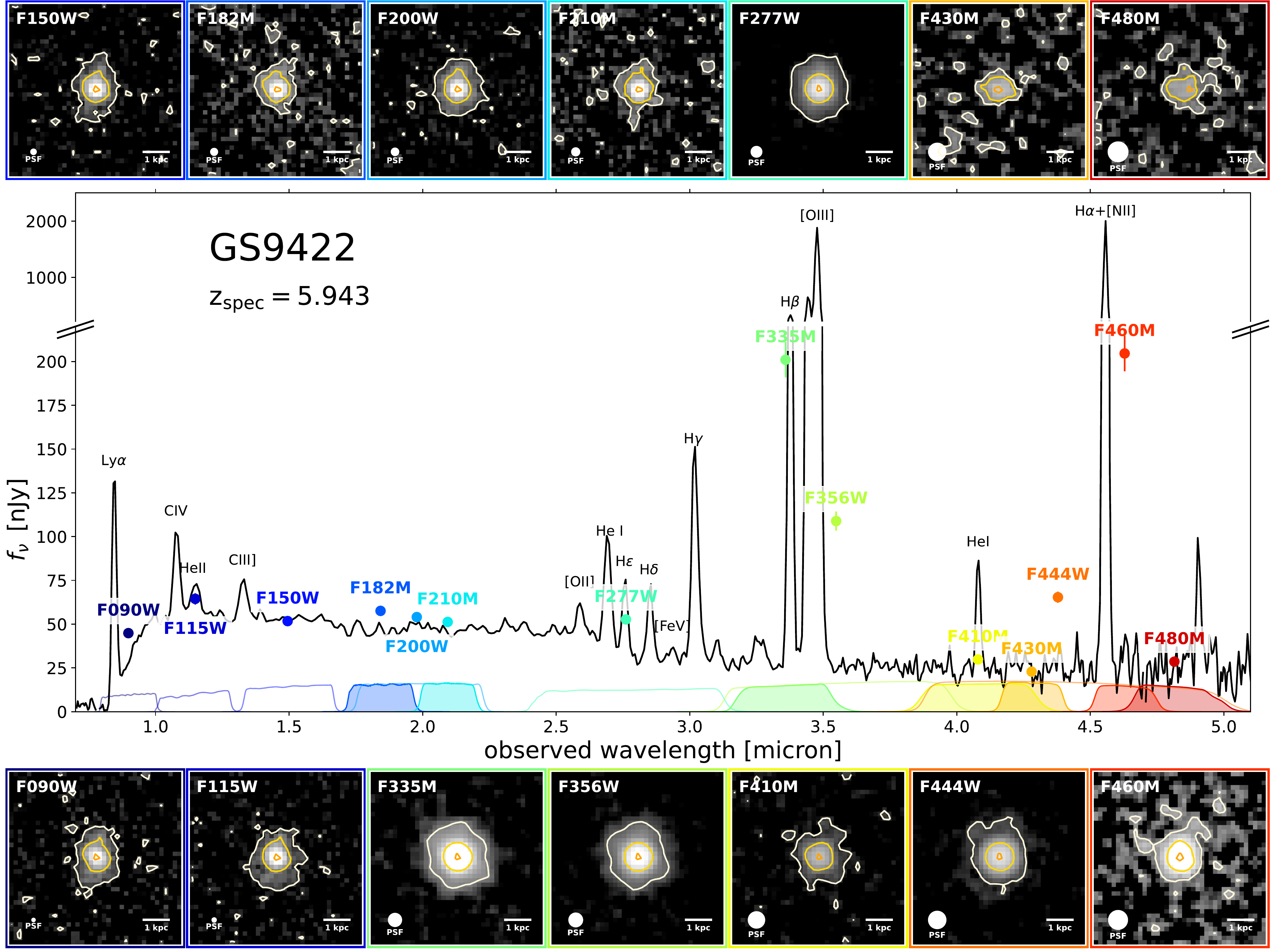}
    \caption{JWST/NIRSpec prism spectrum and the JWST/NIRCam 14-band photometry from JADES and JEMS. The photometry traces the spectrum well: several filters (e.g., F335M, F356W, F444W, F460M) have a significant contribution from emission lines. The transmission curves of the NIRCam filters are plotted at the bottom. The shaded transmission curves indicate the 7 medium bands. The stamps are ordered by wavelength from left to right. The contours indicate the $90^{\rm th}$, $98^{\rm th}$, and $99.9^{\rm th}$ percentiles of the flux.  The filters that have a significant contribution of emission lines are shown at the bottom. We find clear morphological differences for the filters that are dominated by the emission lines to the ones dominated by the continuum, while for the latter ones, we also find a wavelength dependence (rest-frame UV versus optical bands). }
    \label{fig:sed}
\end{figure*}

\begin{table}
	\centering
	\caption{Kron aperture photometry of the 14 bands from JADES and JEMS. The table lists the filters, exposure times, fluxes (in nJy), and key emission lines traced by the different filters. The Kron radius is 1.23\arcsec.}
	\label{tab:photometry}
	\begin{tabular}{lccl} % four columns, alignment for each
		\hline
		Filter & Exposure time & Flux & Key emission lines \\
		  & [hr] & [nJy] &  \\
		\hline
    F090W & 19.9 &  $44.9 \pm 1.1$ & Ly$\alpha$ \\ 
    F115W & 33.6 &  $64.5 \pm 0.8$ & \CIV, \HeII \\ 
    F150W & 19.9 &  $51.7 \pm 0.9$ & \CIII \\ 
    F182M & 7.2 &  $57.6 \pm 1.9$ & none \\ 
    F200W & 13.7 &  $54.0 \pm 0.9$ & none \\ 
    F210M & 7.5 &  $51.3 \pm 2.1$ & none \\ 
    F277W & 19.8 &  $52.7 \pm 0.3$ & \OII, H$\gamma$, H$\delta$, H$\varepsilon$, \HeI \\ 
    F335M & 13.7 &  $201.0 \pm 0.5$ & \OIII, H$\beta$ \\ 
    F356W & 13.7 &  $108.9 \pm 0.3$ & \OIII, H$\beta$ \\ 
    F410M & 19.8 &  $29.8 \pm 0.5$ & \HeI \\ 
    F430M & 1.9 &  $22.9 \pm 2.4$ & none \\ 
    F444W & 20.2 &  $65.4 \pm 0.4$ & H$\alpha$, \NII \\ 
    F460M & 1.9 &  $204.6 \pm 3.6$ & H$\alpha$, \NII \\ 
    F480M & 3.9 &  $28.6 \pm 2.4$ & none \\ 
		\hline
	\end{tabular}
\end{table}

\subsection{GS9422 in the HUDF}

GS9422 ($\mathrm{RA}=53.121757^{\circ}$ and $\mathrm{Dec}=-27.797638^{\circ}$) is mentioned the first time in the literature by \citet{coe06}, who performed a Bayesian photometric redshift analysis in the Hubble Ultra Deep Field \citep[HUDF;][]{beckwith06}. They classified GS9422 as a high-redshift galaxy candidate with a photometric redshift of $z\sim1.1$. In 2015, with better photometry thanks to Hubble Space Telescope's Wide Field Camera 3 (WFC3), GS9422 has been firmly put at $z>5$, with \citet{finkelstein15} and \citet{bouwens15} determining a photometric redshift of 5.89 and 5.76, respectively. Their photometric redshift estimates are close to the spectroscopic redshift of $z_{\rm spec}=5.943$ obtained with NIRSpec \citep{bunker24} and NIRCam grism observations.

\subsection{JWST NIRCam and NIRSpec datasets}

Lying in the HUDF region and being a high-redshift galaxy candidate, GS9422 has been extensively observed with JWST. In particular, the \textit{JWST Advanced Deep Extragalactic Survey} (JADES; PIs Rieke and L\"utzgendorf; \citealt{eisenstein23_jades}) is a Guaranteed Time Observation (GTO) programme of the NIRCam and NIRSpec instrument teams with a goal to build synergies between imaging and spectroscopy. JADES provides imaging of an area of $\sim175$ arcmin$^2$ in the GOODS-S and GOODS-N extragalactic fields with an average exposure time of 20 hr spread over $8-10$ filters. The NIRCam observations are complemented with NIRSpec multi-object spectroscopy of over 5000 HST and JWST-detected sources with 5 low, medium, and high-resolution dispersers covering $0.6-5.3~\micron$. 

GS9422 was observed in the JADES survey with all five spectral modes, receiving 28 hr integration in Prism/CLEAR and 7 hr integration in each of G140M/F070LP, G235M/F170LP, G395M/F290LP and G395H/F290LP. In addition, GS9422 was imaged (NIRCam ID 113585) in the following 9 JWST/NIRCam filters as part of JADES: F090W (20 hr integration), F115W (34 hr), F150W (20 hr), F200W (14 hr), F277W (20 hr), F335M (14 hr), F356W (14 hr), F410M (20 hr) and F444W (20 hr). Parts of these data have been released \citep{bunker24, eisenstein23_jades, eisenstein23_jof, rieke23}, and we use the reduced spectra released as part of the JADES Public Data Release. The spectroscopic data are the focus of \citet{cameron24}, while we expand upon that analysis by including also the imaging data. 

The RGB colour composite (R=F444W, G=F200W, and B=F090W) of GS9422 is shown in Figure~\ref{fig:slit_position}. The galaxy is indeed compact, though we see hints of a colour gradient, with a green-bluish extension in the northwards direction. We also overplot the slit NIRSpec-MSA shutters, which shows that the NIRSpec spectrum captures the majority of the galaxy's emission. The path-loss corrections (including slit losses) are estimated assuming a point-source geometry, as described in \citet{bunker24}. The assumption of the point-source geometry is expected to hold well given the compact nature of GS9422 (Fig.~\ref{fig:slit_position} and Section~\ref{sec:morphology}).

Crucially, we include in the analysis here the observations from the survey \textit{JWST Extragalactic Medium-band Survey} (JEMS; PIs Williams, Maseda \& Tacchella; \citealt{williams23_jems}). JEMS is the first public medium-band imaging survey carried out using JWST/NIRCam. These observations use the following $\sim2$ and $\sim4~\mu$m medium-band NIRCam filters: F182M, F210M, F430M, F460M, F480M, with exposure times of 3.9 hr, 3.9 hr, 1.9 hr, 1.9 hr and 3.9 hr, respectively. For GS9422 at $z=5.943$, the medium-band filter F460M probes the $\mathrm{H}\alpha+$\NII emission line complex, while the medium-bands blue-ward (F430M) and red-ward (F480M) probe the underlying rest-frame optical continuum (Figure~\ref{fig:sed}). As we see in Section~\ref{sec:morphology}, this is key to constraining the difference in morphological structure of the emission lines with respect to the underlying continuum, highlighting the power of medium-band imaging with JWST. 

We also note that this source is covered by a full suite of MIRI filters from $5-21~\mu$m from the SMILES program \citep[PI Rieke; Alberts et al. in preparation;][]{lyu24}. Our source is not detected in any of the filters, although this is not surprising owing to the relatively shallow depth of the data, the faintness of the continuum, and the blue colour of our source. 

Finally, we make use of the survey \textit{First Reionization Epoch Spectroscopically Complete Observations} (FRESCO; PI Oesch; \citealt{oesch23}). While GS9422 is clearly detected in the F444W grism observations, we do not include this because the grism spectrum is shallower and lower resolution than the available NIRSpec spectra. Importantly, the grism spectrum is fully consistent with the NIRSpec one. We use the two medium-band images in the short-wavelength filters (F182M and F210M), each of which adds an additional 3.3 hr of depth in those filters.

In summary, by combining data from JADES, JEMS and FRESCO, GS9422 has imaging in 14 different NIRCam filters and spectroscopy with 5 different spectral modes. Figure~\ref{fig:sed} shows the NIRSpec prism spectrum together with the medium- and wide-band NIRCam photometry (Table~\ref{tab:photometry}). The NIRSpec spectrum puts this galaxy firmly at a redshift of $z_{\rm spec}=5.943$. In the rest-frame optical, we find that the galaxy has strong emission lines, which imprint themselves onto several medium- and wide-band filters. Specifically, F335M is strongly affected by \OIII and H$\beta$, while F460M is dominated by H$\alpha$ and \NII. Importantly, the two medium bands F430M and F480M are free of strong emission lines, probing the rest-frame optical continuum. F277W, F356W, F410M, and F444W do contain both a mixture of emission lines and continuum emission. The rest-frame UV continuum is probed by the F182M, F200W, and F210M filters, while the F090W, F115W, and F150W have contributions from Ly$\alpha$, \CIII, \CIV~ and \HeII. Specifically, Ly-$\alpha$ contributes 32\% of the flux to the F090W band. Furthermore, Figure~\ref{fig:sed} plots the cutouts of the 14 bands, which show distinct morphological features, which we quantify and discuss in Section~\ref{sec:morphology}.

\subsection{Basic properties of GS9422}

GS9422 has a UV magnitude of $M_{\rm UV}=-19.45\pm0.17$ mag, which means that it is about 1.5 mag fainter than the knee of the Schechter UV luminosity function at $z\approx6$ ($M_{\rm UV}^{*} \approx -21.0$; \citealt{bouwens15}). We estimate the rest-frame equivalent width for the H$\alpha$+\NII emission line complex from the medium-band photometry with 

\begin{equation}
    \mathrm{EW(H\alpha+[NII])} = \frac{(\mathrm{F460M}-C)\times\mathrm{BW}}{C\times(1+z)},
\label{eq:ew}
\end{equation}
where BW is the bandwidth of the F460M filter ($0.228~\micron$) and $C$ is the flux density of the continuum, which we estimate by taking the average of F430M and F480M. This gives a rest-frame equivalent width of EW(H$\alpha$+\NII) $=2284_{-160}^{+180}$ \AA, which is consistent within the uncertainties (dominated by estimating the continuum flux level) with the direct estimate from the prism spectrum, which is $1566\pm110$ \AA\ (when estimating the continuum from the prism itself) and $2254\pm54$ \AA\ (when estimating the continuum from F444W; \citealt{boyett24}). Similarly, we obtain a \OIII+H$\beta$ equivalent width for GS9422 of EW(\OIII+H$\beta$) $=3420_{-242}^{+265}$ \AA\ from photometry, while the prism spectrum gives $2923\pm57$ \AA\ (when estimating the continuum from the prism itself) and $3290\pm73$ \AA\ (when estimating the continuum from F444W). These are high equivalent widths, but not unseen at these redshifts. GS9422 lies at the upper envelope of the equivalent width distribution at $z\approx6$ \citep{smit15, endsley24, matthee23_eiger}. 

As we discuss below, since large parts of the spectrum and photometry are dominated by nebular emission (nebular emission lines and nebular continuum), it is difficult to infer the stellar mass and stellar population parameters of the underlying galaxy. Assuming standard SED modelling (stellar emission, processed by gas and dust), \citet{simmonds24} used \texttt{Prospector} \citep{johnson21} to infer for GS9422 a stellar mass of $M_{\star}=10^{9}~M_{\odot}$, a star-formation rate of  $\mathrm{SFR}=40~M_{\odot}/\mathrm{yr}$, a stellar age of $t_{50}=13$ Myr and a significant attenuation with $\tau_{\rm V}\approx1$. On the other hand, \citet{scholtz23} used \texttt{Beagle} \citep{chevallard16} to infer a stellar mass of $10^{7.7}~M_{\odot}$ and a star-formation rate of $\mathrm{SFR}=5.4\pm0.1~M_{\odot}/\mathrm{yr}$. This shows that the uncertainties are large, with roughly an order of magnitude when it comes to the stellar mass. This large difference in the stellar mass comes from the assumed star-formation history (SFH; and the associated dust attenuation): \citet{simmonds24} assume a ``non-parametic'' SFH, while \citet{scholtz23} assume a ``delayed exponential+burst'' SFH. The non-parametric SFH allows hiding a lot of mass in older phases of star formation, outshone by more recent stars \citep{whitler23_sfh, tacchella22_highz, tacchella23_metal}. 

\citet{helton24_overdensity2} reported the finding of $17$ high-redshift galaxy overdensities in GOODS-N and GOODS-S with JWST/NIRCam, using imaging from JADES and JEMS alongside wide field slitless spectroscopy from FRESCO. One of these overdensities, JADES$-$GS$-$OD$-5.928$, contains $14$ galaxies with $\mathrm{S/N} > 3$ detections of $\mathrm{H}\alpha$, including GS9422. This protocluster candidate (with a mean location of $\mathrm{R.A.} = 53.13527$, $\mathrm{Dec.} = -27.78893$) is centred $\approx 0.89$ arcminutes ($\approx 2.2$ cMpc) away from GS9422 and represents an overdensity around $5.9 \pm 0.8$ times that of a random volume. Furthermore, \citet{Witstok24} found that roughly a third of their LAEs coincided with large-scale galaxy overdensities in GOODS-S at $z \approx 5.9$, potentially corresponding to ionised bubbles with sizes of $R_{\mathrm{ion}} \approx 3.6-6.4$ cMpc (roughly corresponding to half of the physical diameter of JADES$-$GS$-$OD$-5.928$, which is $\approx 9.6$ cMpc). This is consistent with the idea that once Ly-$\alpha$ escapes GS9422, it will be able to propagate towards us and is not absorbed by the neutral intergalactic medium.

\subsection{Photometric confirmation of the Balmer jump}

\begin{figure*}
	\includegraphics[width=\textwidth]{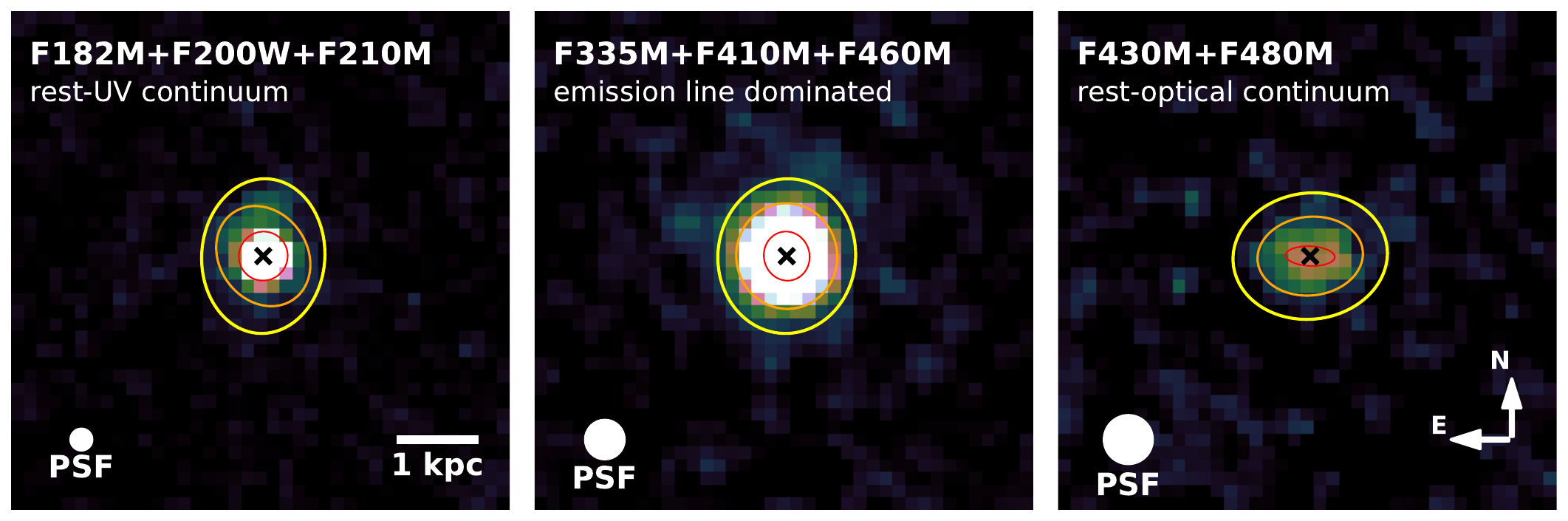}
    \caption{Stack of the rest-UV continuum-dominated bands (F182M, F200W and F210M; left panel), the emission-line-dominated bands (F335M, F410M and F460M; middle panel), and the rest-optical continuum-dominated bands (F430M and F480M; right panel). The ellipses in each panel are from the elliptical isophotal analysis (Fig.~\ref{fig:isophotal}) and mark the ellipses with the semi-major axis lengths of 0.06, 0.12 and 0.18 arcsec (corresponding to 0.35, 0.71 and 1.1 kpc, respectively). The circle in the bottom left indicates the size of the PSF of the longest-wavelength band of the stack. The rest-optical continuum emission has a different spatial orientation and extent than the rest-UV continuum and line emission. Consistent with this, we find the EW($\mathrm{H\alpha}+[\mathrm{NII}]$) to decrease from the centre of the galaxy outwards along the E-W direction, while it increases along the N-S direction (Fig.~\ref{fig:ew_map}).}
    \label{fig:stack}
\end{figure*}

\begin{figure*}
	\includegraphics[width=\textwidth]{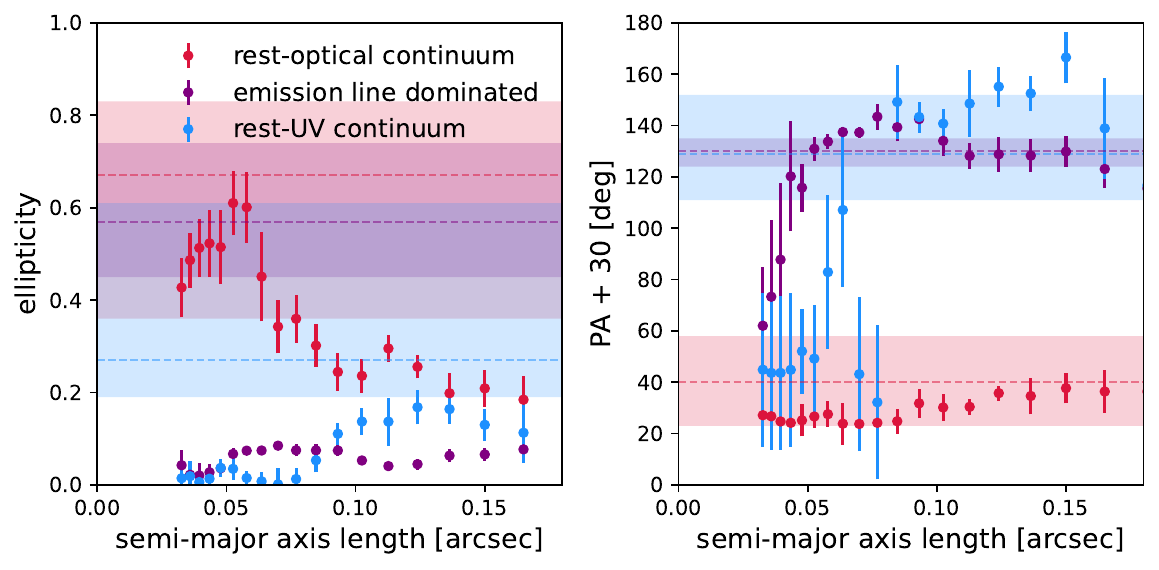}
    \caption{Elliptical isophotal analysis of the stacked images (Fig.~\ref{fig:stack}) that probe the rest-UV continuum emission (F182M, F200W and F210M; blue), emission lines (F335M, F410M and F460M; purple), and rest-optical continuum emission (F430M and F480M; red). We plot the ellipticity ($\varepsilon=1-b/a$; left panel) and position angle (PA; right panel) as a function of semi-major axis length. We have added 30$^{\circ}$ to the PA for clarity. The rest-optical continuum emission ($\varepsilon=0.30_{-0.07}^{+0.18}$) shows a factor of 3 larger ellipticity than the emission line dominated light ($\varepsilon=0.07_{-0.02}^{+0.01}$). The orientation (i.e. PA) of the ellipses are rotated from each other by $\sim90^{\circ}$. The rest-UV and line emission only show minor differences. The horizontal lines and shaded regions indicate the best-fit values and $16^{\rm th}-84^{\rm th}$ percentiles from the \texttt{pysersic} analysis that takes into account the PSFs (Table~\ref{tab:bestfitmorpho}). We find overall good agreement between the isophotal analysis and the profile-forward modelling approach: as expected, since the PSF circularises the appearance of galaxies, the isophotal approach leads to lower ellipticities than the forward-modelling approach.}
    \label{fig:isophotal}
\end{figure*}

The prism spectrum clearly shows a spectral discontinuity at an observed wavelength of $\sim2.530$ \micron~ in the continuum, which corresponds to the Balmer jump at rest-frame wavelength of 3645 \AA. \citet{cameron24} estimated the Balmer jump to be $15.0 \pm 0.9$ nJy. 

Estimating the Balmer jump strength from the photometry alone is challenging because of the contamination of emission lines. Specifically, F277W, F335M, F356W, F410M, F444W, and F460M do all trace emission lines and are therefore not suited for estimating the continuum. We use F182M, F200W and F210M to estimate the UV continuum slope by fitting a simple power-law to those bands and extrapolate this best-fit to $\lambda_{\rm rest}=3645$ \AA. Redwards of the Balmer jump, we use F430M and F480M to estimate the continuum level, assuming a simple constant in $f_{\nu}$. Propagating the uncertainties gives an estimate of the Balmer jump strength of $17.4 \pm 1.9$ nJy, which is consistent with the estimate from the prism spectrum within the uncertainties. We also estimate the Balmer jump as a fraction, finding for F(4200\AA)/F(3500~\AA) a value of $0.57 \pm 0.04$.

\section{Morphology: difference between emission lines and continuum}
\label{sec:morphology}

We assess the morphological structure of GS9422. While GS9422 is compact (Fig.~\ref{fig:slit_position}), we find that the morphology changes as a function of wavelength (Fig.~\ref{fig:sed}). In this section, we perform an isophotal analysis (Section~\ref{subsec:isophotal}) and parametric modelling (Section~\ref{subsec:morph_param}) in order to assess the morphology of GS9422 in the rest-frame UV continuum, in the rest-frame optical continuum, and of the rest-optical emission lines. The main goal is to understand whether these emissions are co-spatial or not. As we show below, there are clear morphological differences, which challenge the interpretation that the full spectrum is dominated by nebular emission.

\subsection{Isophotal analysis}
\label{subsec:isophotal}

We start with a non-parametric morphological analysis, where we directly measure the light distribution on images. We perform a simple average stack to trace the rest-UV continuum emission (stack of F182M, F200W, and F210M), rest-optical emission lines (stack of F335M, F410M, and F460M), and rest-optical continuum emission (stack of F430M and F480M). We show those stacks in Figure~\ref{fig:stack}. Already by eye, we can see that the rest-UV continuum stack and the emission line dominated stack both show a rather circular light distribution (possibly slightly elongated in N-S direction), while the rest-optical continuum shows a clear elongation in the E-W direction. 

To quantify this further, we fit elliptical isophotes to those three stacks using the iterative method described by \citet{jedrzejewski87} as part of \texttt{astropy}. The centre was determined from the F277W image and held fixed during the fitting. The best-fit ellipses are shown in Figure~\ref{fig:stack}. While the rest-UV continuum stack and the emission line dominated stack show a similar ellipticity ($\varepsilon=1-b/a$, where $a$ and $b$ are semi-major and semi-minor axis lengths) and position angle (PA), the rest-optical has a higher ellipticity and a PA that is $\sim90^{\circ}$ rotated. 

Figure~\ref{fig:isophotal} plots the ellipticity (left panel) and PA (right panel; plotted with $+30^{\circ}$ for clarity) as a function of the semi-major axis length. The ellipticity is $\varepsilon=0.3-0.6$ for the rest-optical emission in the central $0.03-0.1$ arcsec ($0.18-0.59$ kpc), which is significantly larger than what we find for the rest-UV and emission line dominated stacks ($\varepsilon<0.1$). The fact that the optical continuum has such high ellipticity is consistent with the idea that it is tracing an edge-on stellar disc. Those differences persist beyond $0.6$ kpc, though they are less significant. We also confirm the stark difference in the orientation of the ellipses: while the rest-optical stack is consistent with a PA of $0^{\circ}$ (E-W direction), ellipses of both the rest-UV and emission lines stacks have a PA of $\sim 100^{\circ}$ (N-S direction).

Since the rest-optical continuum and the emission lines are probed by medium-band filters in a similar wavelength range, this indicates that the different morphologies and orientations are intrinsic and largely independent of PSF variations. In particular, F460M (probing H$\alpha$+[NII]) has a very similar PSF as F430M and F480M used to probe the nearby optical continuum. Also when considering PSF effects (i.e. fitting a PSF-convolved S\'{e}rsic model with \texttt{pysersic} to those images; see below for details), we find consistent results regarding ellipticities and PAs. In Figure~\ref{fig:isophotal}, the best-fit and $16^{\rm th}-84^{\rm th}$ ellipticity and PA from \texttt{pysersic} are shown as dashed lines and shaded regions, respectively. As expected, the ellipticities increase when correcting for the PSF (i.e. the PSF circularises images), but we still find a larger ellipticity for the rest-optical continuum emission than for the rest-UV and the emission line dominated emission.

\begin{figure}
	\includegraphics[width=\linewidth]{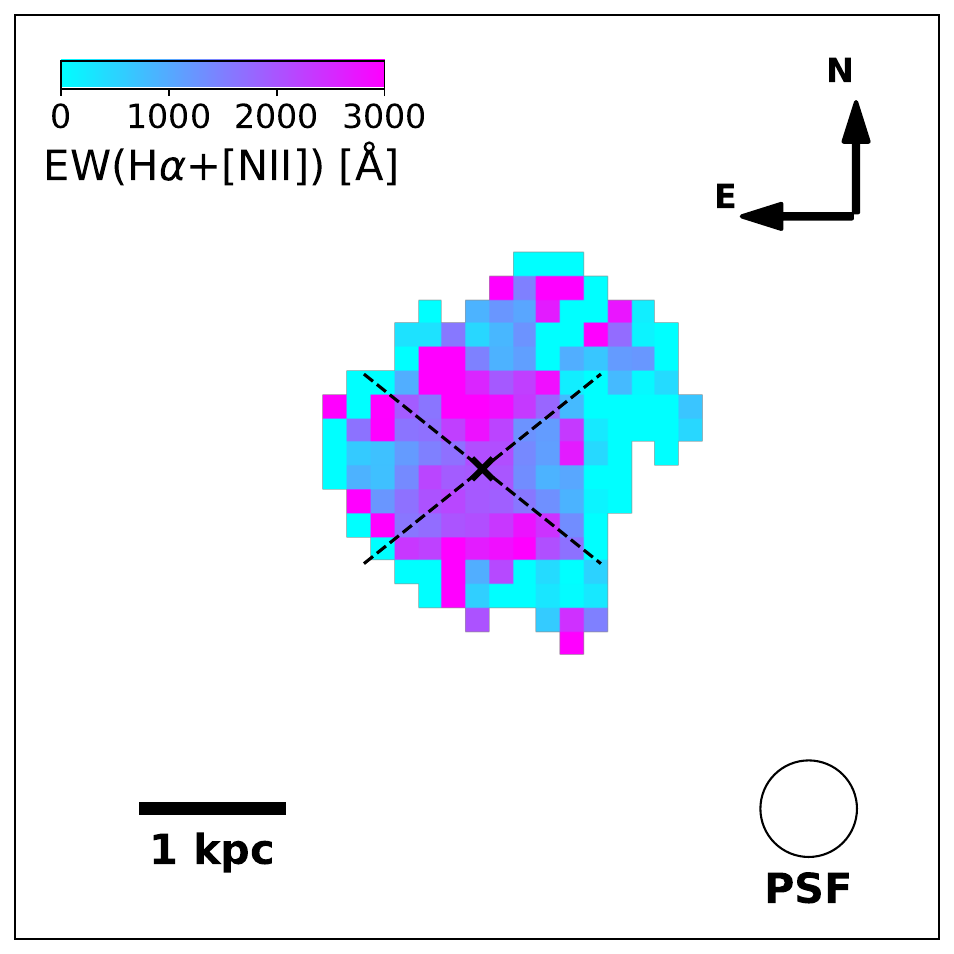}
    \caption{$\mathrm{H\alpha}+[\mathrm{NII}]$ equivalent width (EW) map, estimated by using Eq.~\ref{eq:ew} on a pixel-by-pixel basis. We find that the EW($\mathrm{H\alpha}+[\mathrm{NII}]$) decreases from the centre of the galaxy outwards along the E-W direction, while it increases along the N-S direction. This is consistent with a cone-like structure that is indicated with the dashed lines.}
    \label{fig:ew_map}
\end{figure}

Given these differences in the emission line dominated bands and the continuum dominated bands, we expect to find a gradient in EW across GS9422. Figure~\ref{fig:ew_map} indeed confirms an $\mathrm{EW(H\alpha+[NII])}$ gradient across GS9422. We obtain $\mathrm{EW(H\alpha+[NII])}$ using our JEMS medium bands (Eq.~\ref{eq:ew}). The central region has an EW of $\sim2000$ \AA, which is consistent with the one from the integrated photometry (EW(H$\alpha$+\NII) $=2284_{-160}^{+180}$ \AA). We find that the EW decreases towards the east and west directions, while it increases towards the north and south directions. We highlight these cone-like behaviours with dashed lines. Consistently, we confirm EW gradients using the NIRSpec data \citep{tripodi24}, though these gradients are rather weak because the shutters are aligned along the WNW direction (Fig.~\ref{fig:slit_position}); this way, the opposite radial trends of EW in the N-S and E-W direction are blended and effectively washed out. Importantly, the EW map in Fig.~\ref{fig:ew_map} is not corrected for PSF effects. However, our results are not due to PSF effects because F430M, F460M, and F480M all have very similar PSFs. On the contrary, PSF effects mean even sharper and more pronounced intrinsic EW gradients.

\subsection{Parametric modelling}
\label{subsec:morph_param}

\begin{figure*}
	\includegraphics[width=\textwidth]{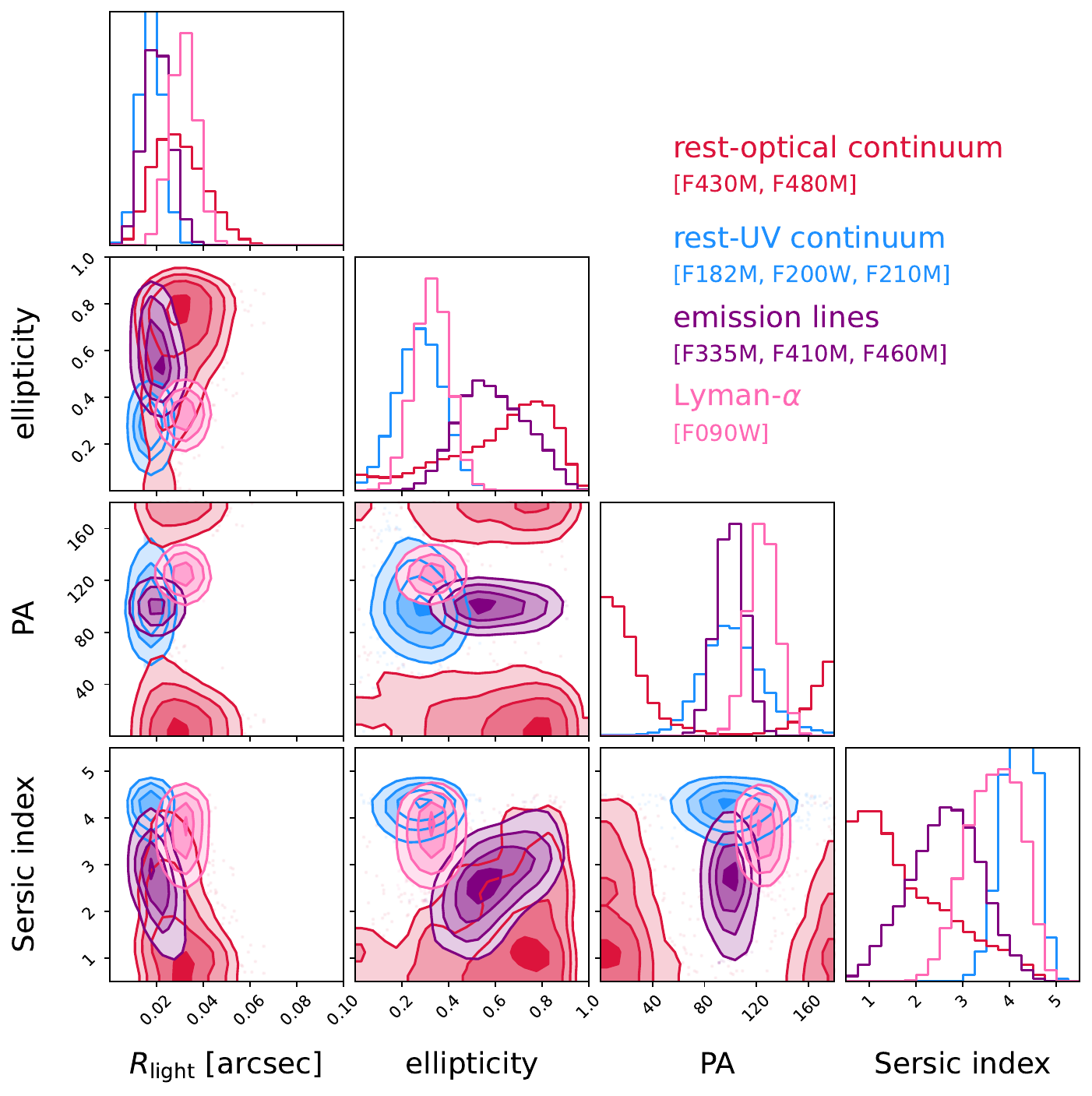}
    \caption{Posterior distributions of half-light size ($R_{\rm light}$), ellipticity ($\varepsilon=1-b/a$), position angle (PA), and S\'{e}rsic index obtained from the morphological analysis with \texttt{pysersic}. The red, blue, purple, and pink contours show the posteriors for the rest-optical continuum emission, rest-UV continuum emission, emission lines (incl. H$\alpha$, H$\beta$ and \OIII), and Ly$\alpha$ emission. While the extent (i.e. half-light sizes) of these different emission tracers are similar with about 0.03 arcsec (0.18 kpc), there are clear differences regarding the shape (ellipticity, PA, and S\'{e}rsic index). Specifically, the rest-optical emission shows a disc-like morphology with a high ellipticity and at $\sim90^{\circ}$ relative to the rest-frame UV emission and the emission lines. This is overall consistent with the isophotal analysis presented in Fig.~\ref{fig:isophotal}.}
    \label{fig:pysersic}
\end{figure*}

\begin{table*}
	\centering
	\caption{The best fit morphological parameters, including half-light radius ($R_{\rm light}$), S\'{e}rsic index ($n$), ellipticity ($\varepsilon=1-b/a$), and PA. The half-light radii are in milliarcsec (mas), where 1 mas = 5.9 pc at $z=5.943$. The full posterior distributions of these parameters are shown in Figure~\ref{fig:eml_ratios}, which highlights significant differences between the rest-UV continuum, emission lines and rest-optical continuum for $n$, $\varepsilon$, and PA (the relative consistency implied by the uncertainties in this table stems from looking at 1D marginalised posteriors).}
	\label{tab:bestfitmorpho}
	\begin{tabular}{lcccccc} % four columns, alignment for each
		\hline
		Tracer & Filters & half-light radius $R_{\rm light}$ & half-light radius $R_{\rm light}$ & S\'{e}rsic index $n$ & ellipticity $\varepsilon$ & PA \\
		  &   & [mas] & [pc] &  &  & [degree] \\
		\hline
        Ly-$\alpha$ & F090W & $31.4_{-2.6}^{+2.5}$ & $184_{-15}^{+15}$ & $3.7_{-0.5}^{+0.5}$ & $0.33_{-0.06}^{+0.05}$ & $125_{-6}^{+7}$ \\
        rest-UV continuum & F182M, F200W, F210M & $15.3_{-0.2}^{+0.5}$ & $90_{-1}^{+3}$ & $4.3_{-0.3}^{+0.1}$ & $0.28_{-0.08}^{+0.08}$ & $100_{-19}^{+22}$ \\
        emission lines & F335M, F410M, F460M & $19.7_{-2.6}^{+3.5}$ & $116_{-15}^{+21}$ & $2.6_{-0.8}^{+0.7}$ & $0.58_{-0.12}^{+0.16}$ & $100_{-6}^{+6}$ \\
        rest-optical continuum & F430M, F480M & $28.6_{-9.3}^{+11.5}$ & $168_{-54}^{+67}$ & $1.5_{-0.7}^{+1.4}$ & $0.67_{-0.31}^{+0.16}$ & $10_{-17}^{+18}$ \\
		\hline
	\end{tabular}
\end{table*}

In order to account for the varying PSF in NIRCam (FWHM of 0.033 arcsec in F090W to 0.164 arcsec in F480M), we forward model and fit for the light distribution. We assume a \citet{sersic68}, axis-symmetric light profile. Our fiducial approach for the profile fitting is \texttt{pysersic} \citep{pasha23}, but we also compare our results with \texttt{ForcePho} (Johnson et al. in prep.). Both codes are Bayesian frameworks, allowing us to understand the degeneracies between different shape parameters. 

As in Section~\ref{subsec:isophotal}, we are primarily interested in understanding if different filters, which trace different rest-frame emission features, have different morphologies. Therefore, we fit a S\'{e}rsic profile to the stacks discussed in the previous section. Specifically, we separately fit the Ly-$\alpha$ emission (F090W filter), the rest-UV continuum emission (stack of F182M, F200W and F210M filters), optical emission lines (stack of F335M, F410M and F460M filters), and rest-optical continuum emission (stack of F430M and F480M). Within the \texttt{pysersic} framework, we use the No-U-Turn-Sampling (NUTS) sampler to estimate the posterior distributions and supply model PSFs (mPSFs). These mPSFs are derived following the method of \citet{ji24}, where we inject WebbPSF models \citep{webbpsf} into JWST level-2 images and mosaic them using the same exposure pattern as our JADES and JEMS observations to provide a composite star field. An mPSF for each band (and observing program) is then constructed from these PSF-mosaics.

The best-fit morphological parameters are tabulated in Table~\ref{tab:bestfitmorpho} and their posterior distributions are plotted in Fig.~\ref{fig:pysersic}. We confirm the overall compact nature of GS9422, with half-light sizes ranging from $90-200$ pc for the different wavelength ranges probed. Specifically, we find that Ly-$\alpha$ is the most extended component with $R_{\rm light}=184_{-15}^{+15}$ pc, while the most compact component is the rest-UV continuum with $R_{\rm light}=90_{-1}^{+3}$ pc. The S\'{e}rsic index of both components is consistent with $n=4$. This is consistent with the ubiquitous Ly-$\alpha$ haloes found around $z=3-6$ galaxies \citep{wisotzki18}. Furthermore, we find that the rest-frame optical continuum emission is more extended than the rest-UV and emission lines. However, the profile of the rest-optical component is disc-like with $n=1.5_{-0.7}^{+1.4}$. 

In addition to these differences in the extent of Ly-$\alpha$, rest-UV and rest-optical continuum, and emission lines, we find clear differences in the ellipticity and position angle, which is consistent with the isophotal analysis in Section~\ref{subsec:isophotal}. Specifically, we find a high ellipticity of $\varepsilon=0.67_{-0.31}^{+0.16}$ at a PA of $10^{+18}_{-17}~deg$ for the rest-optical continuum emission, while all other components have a lower ellipticity and a PA of $\sim100^{\circ}$ (specifically $\sim100^{+6}_{-6}$). The PAs of the optical continuum and optical emission lines are orthogonal and inconsistent at several sigmas, further and unambiguously indicating that the optical continuum cannot be nebular dominated. This is overall consistent with the picture obtained from the EW map in Fig.~\ref{fig:ew_map}, where the rest-frame optical continuum emission is a disc-like component in the horizontal direction, while the emission lines are vertically extended. 

The ellipticities inferred from \texttt{pysersic} are overall larger than what we obtained from the isophotal analysis (see Fig.~\ref{fig:isophotal}). This is expected since the isophotal analysis does not correct for the PSF and the PSF circularises the flux distribution, i.e., lowers the ellipticity. Importantly, we find excellent agreement for the PA between \texttt{pysersic} and the isophotal analysis.

We compared our morphological analysis with \texttt{ForcePho} (Johnson et al., in prep.). Specifically, because of the compact nature of this source, \texttt{ForcePho} has the advantage of being able to work on individual exposures, thereby making better use of the sub-pixel information than when working on the mosaics themselves. We find \texttt{ForcePho} half-light sizes are in the range of 11-30 mas, consistent with the ones from \texttt{pysersic}. Within \texttt{ForcePho}, we also tested whether this galaxy warrants a two-component fit (compact core plus extended component; \citealt{tacchella23, baker24}), but those fits did not run successfully, and no evidence for a second component could be found.

%We model the morphological properties of the galaxy as traced by different emissions with the forward-modeling photometry code \textit{Forcepho} (cite Johnson in prep. ???). \textit{Forcepho} uses mixtures of Gaussians to characterize the half-light radius, $R_{50}$ (arcsec), and Sersic index, $n$, of a galaxy’s flux profile simultaneously for customized combinations of filters. This modeling over individual exposures allows for sub-pixel level morphological measurements. 

%A two-component fit characterizing the morphology of the compact core and extended emission is achieved by allowing the model parameters to vary within different bounds for each component, $0.5 < n < 1.5$ and $R_{50} < 0.01$ for the core, and $0.8 < n < 6$ and $0.001< R_{50} < 1$ (default \textit{Forcepho} bounds) for the extended emission. We run \textit{Forcepho} on GS9422 with the following filter combinations corresponding to emission sources:
%\begin{itemize}
%    \item all 14 filters (F090W, F115W, F150W, F182M, F200W, F210M, F277W, F335M, F356W, F410M, F430M, F444W, F460M, F480M)
%    \item rest-UV (F150W, F182M, F200W, F210M or F182M, F200W, F210M)
%    \item Ly$\alpha$ (F090W)
%    \item nebular continuum (F430M, F80M)
%    \item emission lines (F335M, F410M, F460M or F277W, F335M, F356W, F410M, F444W, F460M)
%\end{itemize}

In summary, we find clear differences in the morphologies at wavelengths tracing different emission features. Specifically, while Ly-$\alpha$ is the most extended component, the rest-UV continuum emission is compact. The rest-optical continuum emission is consistent with being an extended edge-on disc. The rest-frame optical emission lines are extended perpendicular to this disc, leading to a cone-like H$\alpha$+\NII equivalent width. In addition to radiative transfer effects that can lead to some differences in the extent of certain emissions (for example Ly-$\alpha$; \citealt{dijkstra06, smith22}), the stark differences between the rest-optical continuum and the optical emission lines suggest that different sources power these emissions. In particular, the different morphologies between the rest-UV continuum, rest-optical continuum, and rest-optical nebular emission lines provide an indication that not all of the continuum is of nebular origin, for which we would expect a more co-spatial distribution of the full emission.

\begin{figure*}
	\includegraphics[width=0.8\paperwidth]{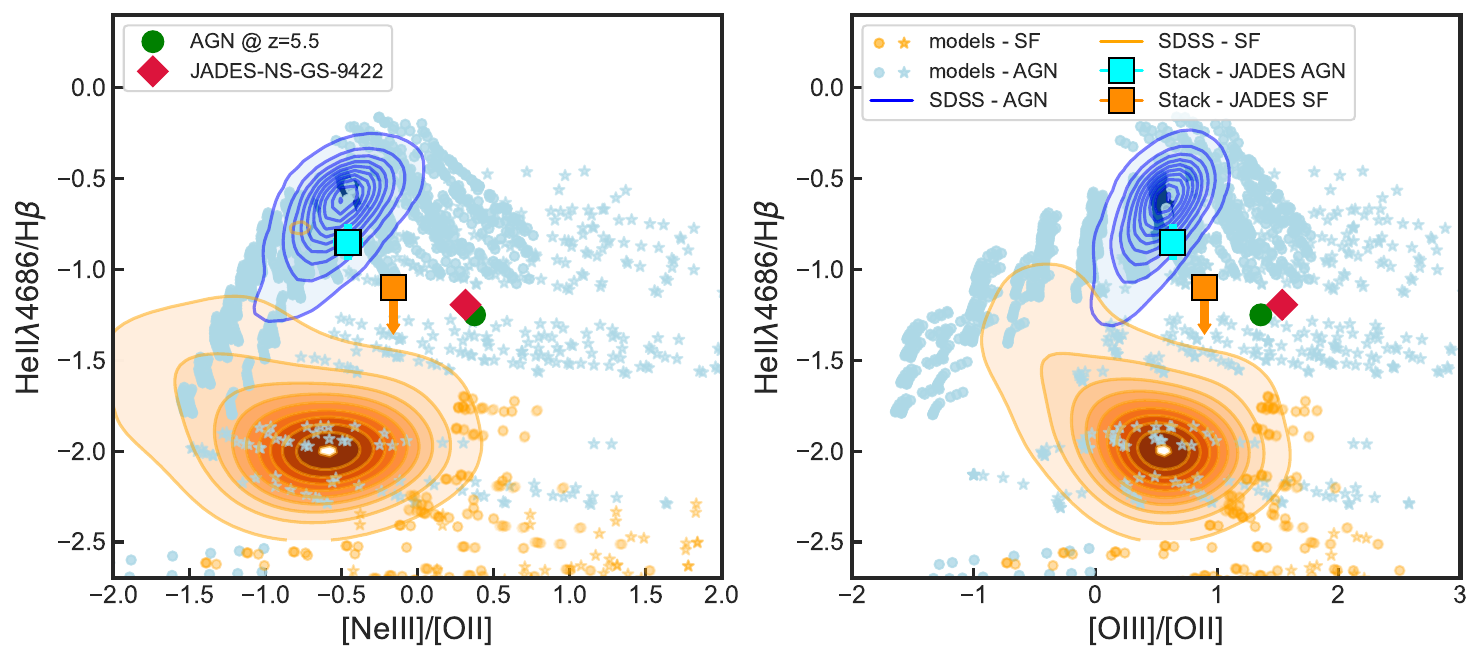}
    \caption{Optical emission line diagnostics diagrams for GS9422. Left: HeII$\lambda4686$/H$\beta$ vs [NeIII]$\lambda3869$/[OII]$\lambda\lambda$3727,29. Right: HeII$\lambda4686$/H$\beta$ vs [OIII]5008/[OII]$\lambda\lambda$3727,29. We show GS9422 as a red diamond, while the green points indicate a well-studied, luminous type-1.8 AGN at $z\sim5.5$ \citep{ubler23}, whose line ratios are fully consistent with GS9422. We compare our observed line ratios to AGN and star-forming galaxies from SDSS (blue and orange shaded contours) and results from the photo-ionisation models for AGN and star-forming galaxies (orange and blue points) from \citet{feltre16, gutkin16, nakajima22}. We also show the AGN and SF stacks from JADES \citep{scholtz23} as cyan and orange squares (upper limit on HeII/H$\beta$ for the SF stack).
    }
    \label{fig:eml_ratios}
\end{figure*}

\begin{figure}
	\includegraphics[width=0.95\columnwidth]{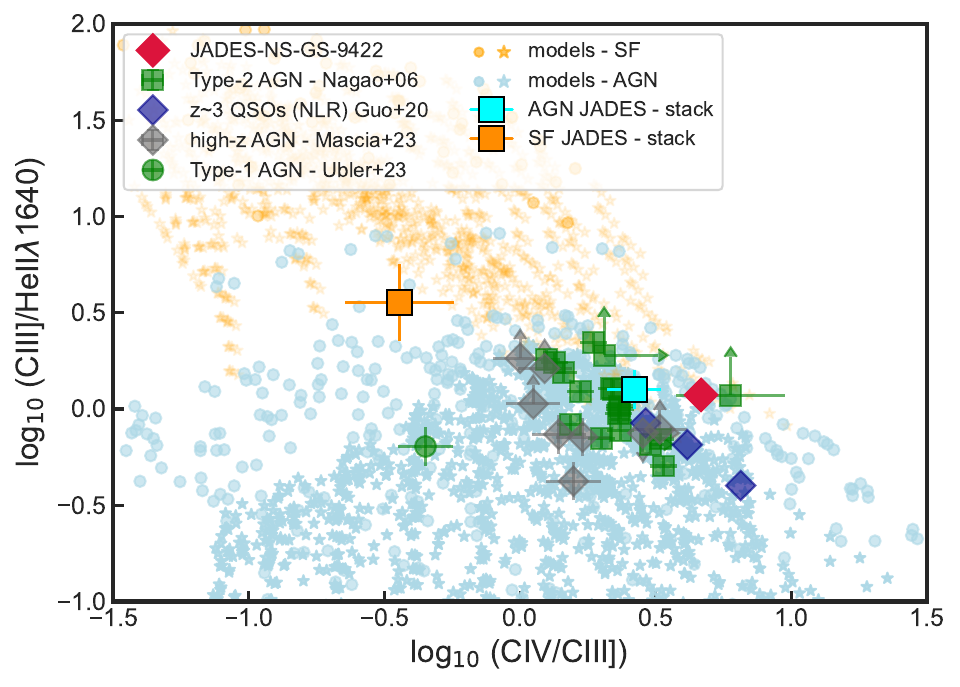}
    \caption{UV emission line diagnostics diagrams using \CIII/\HeIIl1640 vs \CIV/\CIII \ with GS9422 shown as a red diamond. For comparison we show the measurements of the Narrow Line Regions of other AGN from the literature at high redshift: quasars with VLT/MUSE observations (blue diamonds; \citealt{guo20}), AGN from VANDELS survey (grey diamonds; \citealt{mascia23}), compilation of type-2 AGN from  \citealt{nagao06}), type-1 AGN at z$\sim$5.5 (green circle; \citealt{ubler23} and Ji et al. in prep.) and results from the photo-ionisation models for AGN and star-forming galaxies (orange and blue points) from \citet{feltre16, gutkin16, nakajima22}. We also show the AGN and SF stacks from JADES survey \citep{scholtz23} as cyan and dark orange squares.
    }
    \label{fig:UV_diagnostics}
\end{figure}

\section{Emission-line diagnostics}
\label{subsec:el_diag}

\subsection{UV and optical emission lines}

Before performing a detailed modelling of GS9422's spectrum (Section~\ref{sec:sed}), we investigate the ionisation source of this galaxy by studying UV and optical emission-line diagnostic diagrams. Importantly, the NIRSpec MSA spectrum contains light from the entire galaxy, given the compact nature of the galaxy (Figure~\ref{fig:slit_position}). \HeII emission is of particular interest, as this line was interpreted by \citet{cameron24} as a sign of an absence of AGN ionisation. The \HeII$\lambda$1640 is detected at 8.5$\sigma$ in the R1000 NIRSpec data (see App.~\ref{app:HeII}), while \HeII$\lambda$4686 at 5.3$\sigma$ in the PRISM observations (Fig.~\ref{fig:sed}). We estimate the EW$_{\rm rest}$ of the \HeII$\lambda$4686 and \HeII$\lambda$1640 as $17.6\pm3.8$ \AA\ and $6.3\pm0.7$ \AA, respectively. These high EW of the \HeII~ lines of GS9422 are consistent with AGN \citep{nakajima22}, but similar \HeII~ EW can be achieved with X-ray binaries \citep{schaerer19} and binary star models \citep{saxena20}.

In Figure \ref{fig:eml_ratios}, we compare the optical emission line diagnostics (\HeII$\lambda$4686/H$\beta$ vs [NeIII]/[OII] and \HeII$\lambda$4686/H$\beta$ vs [OIII]/[OII]) of our target (red diamond) against local galaxies (SDSS, blue contours are AGN, orange contours are star-forming galaxies), the type-1.8 AGN GS-3073 (green circle) at $z\sim5.5$ (similar to GS9422, and for which we have only taken the narrow line components), which has been thoroughly studied both with JWST and ground-based observations \citep[e.g.][]{vanzella10, grazian20, ubler23}, the stacks from the JADES survey for AGN (cyan square) and the upper limit for SF galaxies (orange square) and photo-ionisation models from \citet{feltre16, gutkin16}, and  \citet{nakajima22}, where blue symbols are for AGN models and orange symbols are for SF models. As shown by \citet{scholtz23}, GS9422 has relatively strong \HeII$\lambda$4686 compared to the H$\beta$ emission, and it is located in the region of the parameter space that is solely occupied by AGN rather than typical star-forming galaxies \citep{feltre16, gutkin16, nakajima22}. Most interestingly, the line ratios observed in GS9422 are identical to those seen in the thoroughly studied AGN GS3073 at $z=5.5$ \citep{ubler23}.

In Figure \ref{fig:UV_diagnostics}, we investigate this galaxy via rest-frame UV emission line diagnostics \citep[e.g.,][]{hirschmann19}. We show GS9422 on the \CIII/\HeIIl1640 vs \CIV/\CIII, as a red diamond, and we compare this galaxy to the models of AGN and star-forming galaxies as blue and orange points, respectively, from \citet{feltre16, gutkin16,nakajima22}, 
with other AGN with deep rest-frame UV spectroscopy from VANDELS survey \citep[grey diamonds;][]{mascia23}, Narrow Line Regions of high-z quasars \citep[blue diamonds;][]{guo20}, and compilation of type-2 AGN and quasars from \citet[][green squares]{nagao06}. We also show again the stacks from the JADES survey for AGN (cyan square) and the upper limit for SF galaxies (orange square).
GS9422's strong \HeIIl1640 emission (with the \CIII/\HeIIl1640 = 0.07) and strong CIV emission (CIV/CIII] = 0.66), make it fully consistent with the AGN photoionisation models and with the Narrow Line Regions of AGN spanning a wide range of redshifts and properties, while it is hard to reconcile with SF models and observations of SF galaxies.

Overall, both rest-frame UV and optical emission line diagnostics show that GS9422 is fully consistent with ionisation by an AGN, without the need for stellar populations with a top-heavy IMF. Consistent with this, \citet{li24} uses \texttt{Cue} \citep{li24_cue}, a flexible neural net emulator for \texttt{CLOUDY}, to infer the shape of the ionising spectrum, and the [O/H], [N/O], [C/O], gas density, and total ionising photon budget directly from the observed emission line fluxes. They find that GS9422 can be well described by a combination of young, metal-poor stars and a low-luminosity AGN, with a young metal-poor stellar population contributing 64\% of the ionising photons and a power-law AGN contributing the remainder.

\subsection{Gas-phase Fe abundance from [Fe\,{\sc v}]$\lambda 4227$}

We note there is another interesting spectral feature at $\lambda_{\rm rest} \approx 4227$ \AA\ ($\lambda_{\rm obs} \approx 2.93~\micron$) detected at $3\sigma$ in the PRISM spectrum as shown in Figure~\ref{fig:sed}. Below (Section~\ref{subsec:pure_agn} and Figure~\ref{fig:sed_model}), the pure AGN model produces an emission line at a similar location but with a lower flux.
The line predicted by the model is [Fe\,{\sc v}]$\lambda 4227$.
The ionisation potential of $\rm Fe^{3+}$ is 54.91 eV, close to the ionisation potential of $\rm He^{+}$, which is 54.42 eV.
Therefore, the presence of [Fe\,{\sc v}]$\lambda 4227$ (requiring the production of $\rm Fe^{4+}$ by ionising $\rm Fe^{3+}$) is not unexpected and, similar to that of \HeII, indicates the ionising spectrum should be hard enough to produce sufficient photons at energies greater than $\sim 54$ eV.
Thus, the observations of [Fe\,{\sc v}]$\lambda 4227$ are fully consistent with AGN as the source of the nebular lines, but we note that [Fe\,{\sc v}]$\lambda 4227$ is also found in some metal-poor blue compact dwarf galaxies in the local Universe \citep[e.g.,][]{thuan05}.
The reason that our model underpredicts this line is likely due to the assumed abundance ratio of Fe/O, which is set to the solar value.

Due to the similar ionisation potentials of $\rm Fe^{3+}$ and $\rm He^{+}$, [Fe\,{\sc v}]$\lambda 4227$ and \HeII$\lambda 4686$ should arise in a similar zone within the gas cloud and we can use their observed flux ratio to estimate the actual abundance of Fe in GS9422.
Assuming the kinematics of [Fe\,{\sc v}]$\lambda 4227$ are tied to those of other optical emission lines bluewards to H$\beta$, we measured its flux to be $(0.76\pm 0.25) \times 10^{-19}~{\rm erg/s/cm^{2}}$ from the PRISM spectrum using \textsc{pPXF} \citep{cappellari04, cappellari17}.
The reason to tie the kinematics of blue optical lines is to mitigate the effect of the difference between the assumed spectral resolution and the actual spectral resolution of the observation, and to avoid slight uncertainties in the wavelength calibration due to the potential offset of the sources within the shutter that, given the uncertainty in the target acquisition, can be up to 0.05$''$.
We checked different tying methods and found that this treatment generally fit the stronger blue optical lines including H$\gamma$ and H$\delta$ better than other configurations.
From the medium resolution spectrum G395M, by tying the kinematics of optical lines and setting the underlying continuum to a flat line to avoid overfitting, we obtained the flux of [Fe\,{\sc v}]$\lambda 4227$ to be $(1.0\pm 0.4) \times 10^{-19}~{\rm erg/s/cm^{2}}$, which is broadly consistent with the measurement in PRISM.

With \textsc{PyNeb} \citep{luridiana15}, we calculate the emissivities for [Fe\,{\sc v}]$\lambda 4227$ and \HeII$\lambda 4686$ assuming a temperature of $T_e = 2\times 10^4$ K and an electron density of $n_e = 10^3~{\rm cm^{-3}}$. We used the CHIANTI atomic data set during the calculation \citep{dere97, young16}.
Combining the emissivities with the observed flux ratio of [Fe\,{\sc v}]$\lambda 4227$/\HeII$\lambda 4686$, we obtained an abundance ratio of $\rm \log(Fe^{4+}/He^{2+}) = -4.31^{+0.15}_{-0.20}$ from PRISM,
for the region where [Fe\,{\sc v}]$\lambda 4227$ and \HeII$\lambda 4686$ both arise.
Considering within the region where He is fully ionised, Fe can have ionisation species other than $\rm Fe^{4+}$, this makes the above estimation a lower limit for the abundance ratio of Fe/He.
Using the abundance ratio of He/H derived by \citet{cameron24}, we obtained $\rm \log(Fe/H) - \log(Fe/H)_\odot \gtrsim -0.77^{+0.15}_{-0.20}$.
From the G395M spectrum, where the noise level is much higher, the constraint becomes $\rm \log(Fe/H) - \log(Fe/H)_\odot \gtrsim -0.61^{+0.19}_{-0.26}$.
We note that the metallicity derived by \citet{cameron24} is $\rm \log(O/H) - \log(O/H)_\odot = -1.10 \pm 0.01$, meaning $\rm \log(Fe/O) - \log(Fe/O)_\odot \sim 0.5$ dex.
We caution that the derivation of the helium abundance by \citet{cameron24} does not include correction for self-absorption of \HeI, which might introduce systematic uncertainties when translating Fe/He into Fe/H using He/H.
Alternatively, if we simply assume the abundance of He as a function of metallicity follows the relation given by \citet{dopita00}, the resultant relative abundance of Fe would be $\rm \log(Fe/O) - \log(Fe/O)_\odot \sim 0.2$ dex.
Even if Fe is not depleted onto dust grains at all (as opposed to the typical depletion that only leaves $\sim 1$ \% of Fe in the gas phase, \citealt{jenkins09}), one still needs a super solar Fe/O to reproduce the observed line fluxes.

In the ISM, Fe is typically enriched by Type-Ia supernovae with a typical delay time of $\sim 10^{8.5} - 10^9$ yr, which is broadly comparable to the age of the Universe at the redshift of GS9422.
In addition, the first Type-Ia supernovae can happen $\sim 3\times 10^7$ yr after the initial star formation \citep[see e.g., Fig.~1 in][]{maiolino19}.
Therefore, the enhanced Fe/O in GS9422 could be a result of recent enrichment by Type-Ia supernovae.
However, the enrichment dominated by Type-Ia supernovae would also produce a significantly subsolar Ne/O \citep{nomoto84, iwamoto99}, which is inconsistent with the near-solar Ne/O derived by \citet{cameron24}.
Therefore, another enrichment mechanism is needed to boost Ne/O.
While it is possible that the Ne/O is maintained by previous enrichment of core-collapse supernovae (CCSNe), whether the Fe/O can remain high with the CCSNe pollution requires more quantitative calculations.
It is also possible that the supersolar Fe/O and near-solar Ne/O are produced by a combination of metal-poor hypernovae and Type Ia supernovae, as speculated for some extremely metal-poor but Fe-enhanced galaxies in the local Universe \citep{kojima21, isobe22, watanabe24}.

\section{SED modelling}
\label{sec:sed}

In this section, we use \texttt{CLOUDY} \citep{ferland98, chatzikos23} modelling to understand the nature of the observed emission. As highlighted in the previous section, the observed emission line ratios of GS9422 are consistent with AGN ionisation (Figs.~\ref{fig:eml_ratios} and \ref{fig:UV_diagnostics}). Nevertheless, given that high ionisation emission can also be caused by a low-metallicity population of massive stars with a normal \citet{salpeter55} IMF, we start out by exploring a pure stellar model (Section~\ref{subsec:pure_stellar}). We then move to a simple, type-2 AGN model that dominates the full spectrum, i.e. the nuclear spectrum and broad lines are absorbed along the line of sight, and we are observing the spectrum of the Narrow Line Region that is fully nebular (Section~\ref{subsec:pure_agn}). Both the pure stellar and the pure AGN model reproduce the spectrum overall well, though it is difficult to explain the different morphologies in continuum and emission lines, including the EW gradients (Section~\ref{sec:morphology}). Additionally, the AGN model struggles to reproduce the low emission line strength of the \OII$\lambda\lambda3726,3729$ doublet. To address these issues, we assess whether we can add stellar emission (represented by an older stellar population) to the AGN model in sufficient amounts to change the morphology, highlighting that this is challenging (Section~\ref{subsec:agn_star}). Finally, in Section~\ref{subsec:agn_star_dla}, we model the AGN emission together with an old and young stellar component and a DLA. This brings the morphological measurements in agreement with the observed spectrum (including the strength of the \OII doublet). 

The SEDs for all stellar components in our modelling are taken as an instantaneous starburst simple stellar population (SSP) from BPASS v2.1 with \citet{salpeter55} IMF with a $300\,\mathrm{M}_\odot$ cutoff and including binaries \citep{eldridge17}. 
The model parameters which we present in this section are not a result of fitting our models to the PRISM spectrum or to emission line fluxes, rather they are physically reasonable values that we have selected to show that each type of model can reproduce the observed spectrum well. 

\subsection{Pure stellar model}
\label{subsec:pure_stellar}

We create our pure stellar model with a spherical geometry within \texttt{CLOUDY} with a constant density. The calculation stops when an electron fraction of 0.01 is reached, or if the temperature falls below 1000 K. In this model, we assume the stars have an age of 1.75\,Myr and the ionisation parameter (i.e., ratio between hydrogen-ionising photons and total hydrogen density; $\log\,\mathrm{U}$) is $-1$. The gas density is set to $\mathrm{n}=10^{5}\,\mathrm{cm}^{-3}$ and the gas-phase metallicity is $\log Z/Z_\odot=-1.0$, with the carbon abundance set separately as $\log Z_\mathrm{C}/Z_{\mathrm{C},\,\odot}=-1.2$. We find that a high gas density allows for more reasonable ionisation parameters to be adopted while still reproducing the observed emission line ratios. We note that this density strongly suppresses the two-photon continuum, however we found that even at lower densities, we were unable to produce a strong turnover. Such high gas densities have been inferred at these redshifts \citep[e.g.,][]{topping24_sf}. 

\begin{figure*}
	\includegraphics[width=\textwidth]{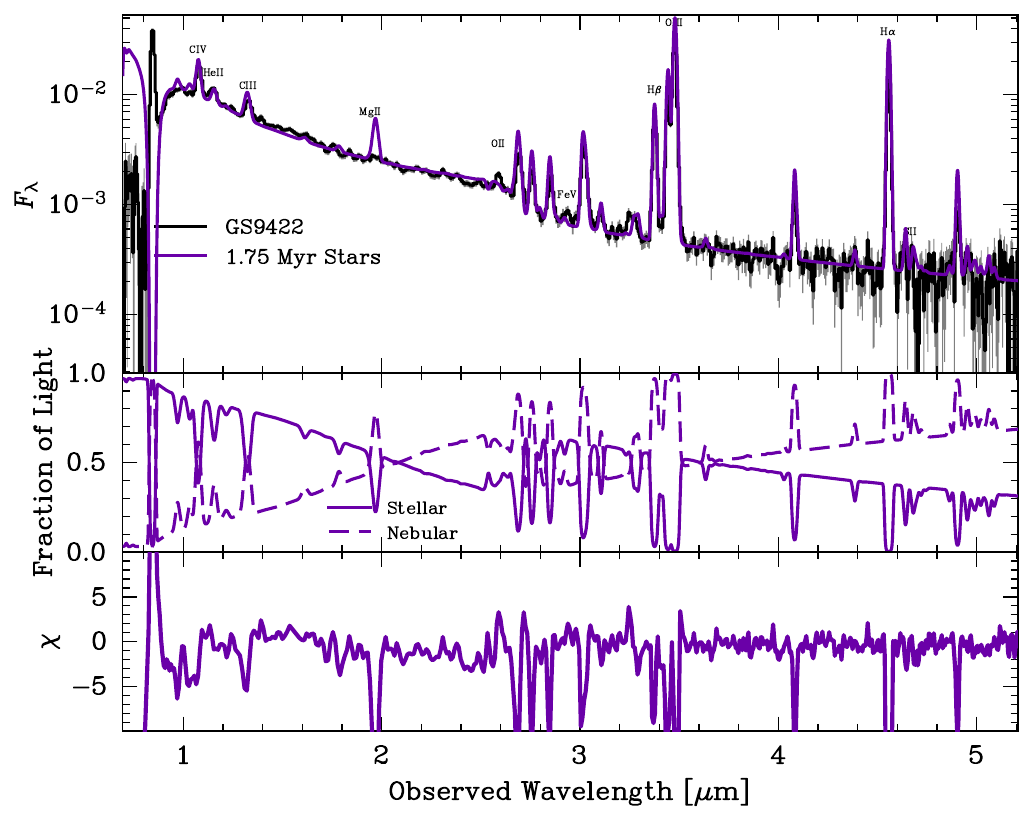}
    \caption{\textit{Top:} the JWST/NIRSpec prism spectrum of GS922 (black) compared to the pure stellar model (purple) with a DLA, which leads to the turnover in the rest-UV. The continuum shape is well reproduced as well as most emission lines. {Mg\,{\sc ii}\,}\,$\lambda \lambda$2796,2803 is not well fit, which may be due to sub-solar {Mg}/{O} abundance or the resonant nature of the line. \textit{Middle:} fractional contribution of nebular (emission lines and continuum) and stellar continuum emission to the model spectrum. The stellar emission dominates in the rest-frame UV, while the nebular emission dominates in the rest-frame optical. \textit{Bottom:} absolute value of $\chi$ between data and model. We find a median $|\chi|$ of 0.95,  which indicates that the model is able to reproduce the data well.}
    \label{fig:sed_model_stellar}
\end{figure*}

\begin{figure*}
	\includegraphics[width=\textwidth]{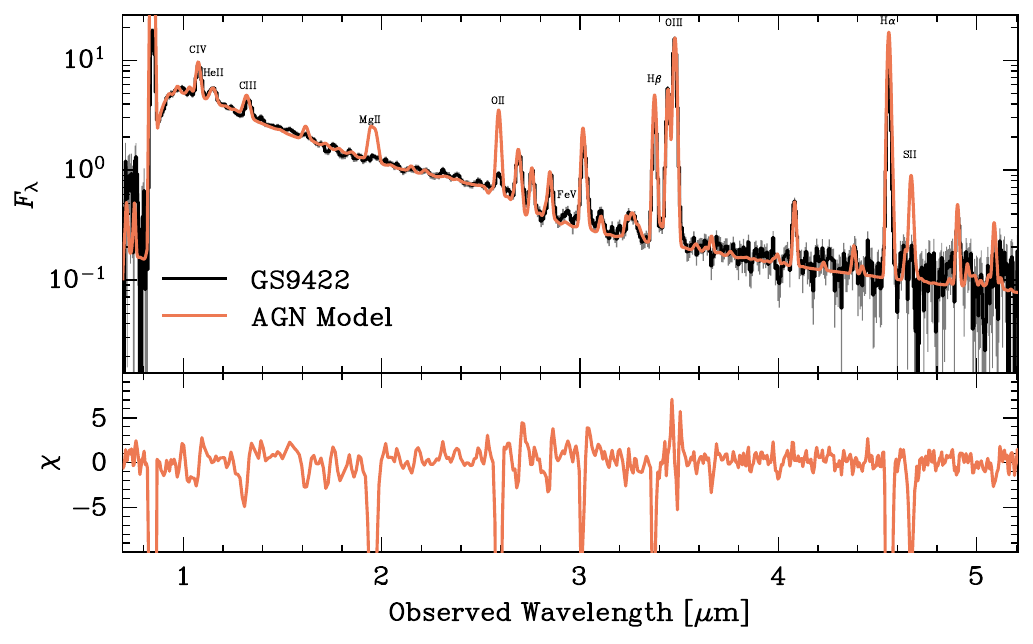}
    \caption{\textit{Top:} the JWST/NIRSpec prism spectrum of GS922 (black) compared to the AGN model (orange). The continuum shape and most of the emission lines are well reproduced. In particular, the \HeIIl1640 and \HeIIl4686 lines are well-fit relative to the nebular continuum. {Mg\,{\sc ii}\,}\,$\lambda \lambda$2796,2803 is poorly fit, which may be due to sub-solar {Mg}/{O} abundance or the resonant nature of the line. \OII$\lambda \lambda 3726,3729$ and [\ion{S}{II}]\,$\lambda \lambda$6716,6731 are poorly fit due to the constant temperature assumption and choice of stopping condition. \textit{Bottom:} absolute value of $\chi$ between data and model. We find a median $|\chi|$ of 1.14,  which indicates that the model is able to reproduce the data well.}
    \label{fig:sed_model}
\end{figure*}

\begin{table}
	\centering
	\caption{The parameters for the \texttt{CLOUDY} AGN model. The gas density and gas temperature define constant gas properties. The overall metallicity is scaled to the oxygen abundance, $\log Z_O/Z_\odot$, except for the carbon abundance, $\log Z_C/Z_\odot$, which is fit separately. The power law slope defines the incident radiation field between >$1\mu \mathrm{m}$ and >$50\mathrm{keV}$ as $f_{\nu}=\nu^{\alpha}$, and the intensity of the incident radiation is set by the ionisation parameter. The gas temperature is well within the estimated electron temperature range from \OIIIl4364~ ($18,300\pm1,500$ K).}
	\label{tab:bestfitcloudy}
	\begin{tabular}{cc} % four columns, alignment for each
		\hline
		Parameter & Value \\
		\hline
        Gas density & $10^{3}\,\textrm{cm}^{-3}$ \\
        Gas temperature & 19,000K \\
        $\log Z/Z_\odot$ & -1.0 \\
        $\log Z_\mathrm{C}/Z_{\mathrm{C},\,\odot}$ & -1.2 \\
        Power law slope & -2.3 \\
        Ionisation parameter & 0.5 \\
		\hline
	\end{tabular}
\end{table}

Figure~\ref{fig:sed_model_stellar} shows the observed PRISM spectrum of GS9422 and the pure stellar model as black and purple lines, respectively. This shows that the pure stellar model is able to describe the observed spectrum well (see $\chi$ values as a function of wavelength in the bottom panel). Notable exceptions are the Ly-$\alpha$ and the \ion{Mg}\,{\sc II}$\lambda\lambda$2796,2803 emission lines. This is not surprising given that both of those lines are resonant lines. The middle panel of Figure~\ref{fig:sed_model_stellar} shows the fractional contribution of stellar and nebular emission of the model. While the rest-UV is dominated by stellar emission, the emission lines and rest-optical are dominated by the nebular emission. 

The UV turnover is reproduced with a DLA. To model the DLA\footnote{\url{https://github.com/joriswitstok/lymana_absorption}}, we use the Voigt approximation of \citet{tasitsiomi06} for the cross-section profile, with a correction to the Voigt profile itself based on quantum-mechanical calculations by \citet{lee13}, and parameterise the DLA profile in frequency rather than wavelength space \citep{webb21}. We obtain a column density of $\mathrm{log}(N_{\mathrm{HI}}/\mathrm{cm^{-2})}=22.9\pm0.1$. Since, in this scenario, the DLA is a local absorber (possibly located at the outer parts of the galactic disc) that uniformly obscures the stellar UV light, the Ly-$\alpha$ emission would likely come from off the disc and is unaffected by the DLA (consistent with its elongated morphology as estimated from F090W), either through recombination emission (the reprocessing of absorbed ionising photons that also contributes to the nebular continuum) or collisionally excited emission. 

We emphasise the good agreement between the observed spectrum and our pure stellar model with a standard, Salpeter IMF. While the two-photon processes powered by exotic stellar populations (IMF that is $10-30\times$ more top-heavy) was the interpretation of GS9422 proposed by \citet{cameron24}, here we show that we can reproduce the UV turnover with a DLA. Nevertheless, there are two reasons why this is not our preferred model. First, this scenario predicts that the emission lines and the rest-optical continuum should have the same morphology. However, as we have shown above (Section~\ref{sec:morphology}), clear morphological differences exist in GS9422. Second, the model assumes that the spectrum arises from solely young stars with an age $1-2$ Myr and a total mass of $\approx10^{7.3}~\mathrm{M}_{\odot}$. While this amount of stellar mass can be formed over such a short timescale (i.e. implied star-formation rate is $\mathrm{SFR}\approx10~\mathrm{M}_{\odot}~\mathrm{yr}^{-1}$), the spatial synchronisation over a scale of 100s of pc is more challenging to explain. Specifically, assuming that the region needs to be physically connected, we can compare this region to the sound speed and the sound crossing time. The sound speed in the typical ISM is 10 km/s ($\sim10$ pc/Myr), but can be lower in higher density gas (0.1 km/s $\approx1$ pc/Myr). This shows that this region is not dynamically connected. This motivates us to look into alternative models, thereby including AGN.

\subsection{Pure AGN model}
\label{subsec:pure_agn}

Since the emission line diagnostic diagrams are consistent with AGN, we now consider an AGN as the powering source for the nebular emission. Specifically, we assume a type-2 AGN, where the source of ionisation (i.e. the accretion disc) and the broad-line region are hidden from our line of sight. We model the AGN emission with a spherical geometry \texttt{CLOUDY} model of constant temperature and density. The gas is illuminated with a power-law SED with three spectral ranges. The low energy (>$1\mu \mathrm{m}$) slope is $f_{\nu}=\nu^{5/2}$, the high energy (>$50~\mathrm{keV}$) slope is $f_{\nu}=\nu^{5/2}$. The mid-range slope is a free parameter of the model, $f_{\nu}=\nu^{\alpha}$ \citep[e.g.,][]{groves04, feltre16}.

Only the diffuse light which is emitted from the gas is observed. All models are iterated to convergence. There are six free parameters, defining the ionising SED (power law slope and ionisation parameter) and the gas properties (temperature, density, metallicity, and carbon abundance).

Defining an appropriate stopping condition can be difficult for a constant temperature model at the very high temperatures which we consider ($\sim20,000$ K). This is because the gas is hot enough to collisionally ionise a significant fraction of hydrogen, causing a breakdown in usual stopping conditions such as electron fraction. However, from the photometric modelling we expect that both the UV continuum and the emission lines are emitted from the compact central source which we are attempting to model. Under this model, both the UV continuum and emission lines are due to nebular emission from the same gas. We can leverage this information to create a stopping condition based on the equivalent width of \HeIIl1640 relative to the nebular continuum, which naturally gives our models the correct strength of nebular emission relative to the (helium) emission lines. This stopping condition tends to lead to density-bounded nebulae emission, which have recently been invoked to explain sub-Case B Balmer ratios, such as those observed in GS9422 \citep{mcclymont24}.

The model parameters are provided in Table~\ref{tab:bestfitcloudy}, and the fit is shown in Figure~\ref{fig:sed_model}. Our model is well fit to the nebular continuum and to most emission lines. However, the fit to \OII\,$\lambda$3727 is notably poor. A high \OII/\OIII ratio generally indicates an ionisation parameter which is too low, however, in this case, it is a side-effect of our modelling assumptions.
This could be alleviated by selecting a different stopping condition which terminates the calculation earlier, however, this would also reduce the strength of the two-photon continuum and therefore sacrifice the quality of our fit to the continuum. We may be able to solve this issue by removing the constant temperature assumption, as this has been shown to cause excessive \OII emission \citep{dors20}. However, it is difficult to achieve sufficiently strong two-photon emission when allowing the gas temperature to be set by thermal equilibrium as there is not enough hot gas.

The over-prediction of [\ion{S}{II}]\,$\lambda \lambda$6716,6731 is likely caused by the same factors as \OII$\lambda \lambda 3726,3729$. $\mathrm{O^{0}}$ and $\mathrm{S^{0}}$ have ionisation energies of 13.6\,eV and 10.4\,eV respectively, which means that $\mathrm{S^{+}}$ abundance generally traces $\mathrm{O^{+}}$ abundance. Therefore, [\ion{S}{II}] emission will be similarly affected by a change in stopping condition or via more refined temperature modelling. However, some of the excess [\ion{S}{II}] emission could be explained by the depletion of sulphur onto dust grains, which is not modelled here \citep{jenkins09}.

Additionally, the Mg\,{\sc ii}\,$\lambda \lambda$2796,2803 doublet is not present in the spectrum of GS9422, however, it often appears in the modelling. As noted in \citet{cameron24}, this is not unexpected as the line is resonant, however, it could also be explained by a sub-solar {Mg}/{O} abundance \citep{kobayashi20}.

\subsection{AGN model with stellar emission}
\label{subsec:agn_star}

\begin{figure*}
	\includegraphics[width=\linewidth]{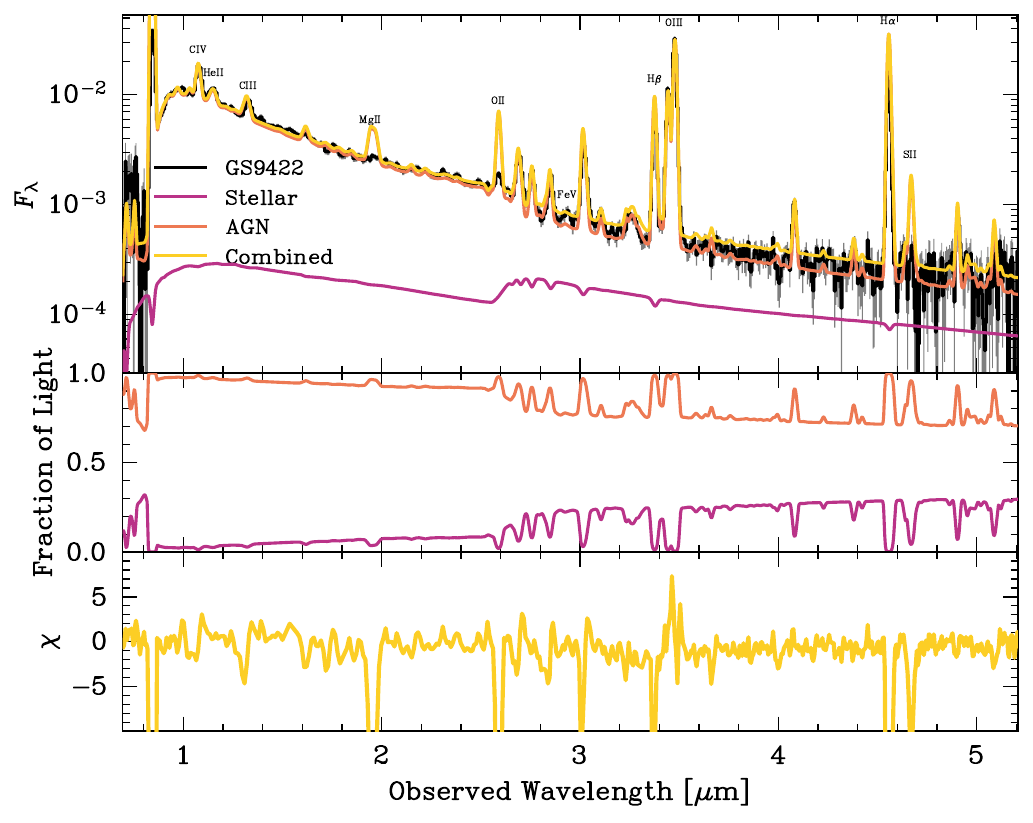}
    \caption{\textit{Top:} the JWST/NIRSpec prism spectrum of GS922 (black) compared to a two-component AGN-stellar model (yellow). The two-component model is made of the AGN model (orange) and the maximum contribution from 50 Myr stellar population obscured by dust (dark magenta). \textit{Middle:} fractional contribution to the emission from the AGN (orange) and the stars (dark magenta). While some flux from stars can be added in the $4-5$ micron range while remaining within reasonable uncertainties, it is insufficient to match the observed morphology. \textit{Bottom:} absolute value of $\chi$ between data and model. We find median $|\chi|$ of 3.46,  which indicates that the model is able to reproduce the data well.}
    \label{fig:combo_sed_model}
\end{figure*}

Given the indication of separate optical and UV/emission line morphologies, it is desirable to create a two-component model, with one component dominating the UV continuum and emission lines, and the other one dominating the optical. We assume that the optical-dominated component is due to a spatially extended stellar emission. The SED for this stellar model is taken to be a 50 Myr old instantaneous starburst. As we are representing a dusty, older stellar population, we apply an SMC dust screen with $\mathrm{A}_{\rm _V}=0.5$ \citep{gordon03}. This is not vital, but it aids the fitting by flattening the spectral slope and allows additional stellar mass to be included in the old stellar component.

By adding this stellar model to our AGN spectrum, we can get an indication of how much flux can be attributed to stellar emission in the rest-frame optical regime. We show in Fig.~\ref{fig:combo_sed_model} that we can add a population of 50 Myr stars to our best-fit AGN model to achieve a $\sim20\%$ stellar contribution to the optical continuum, while still remaining within reasonable uncertainties. However, adding more flux than this causes the optical emission to become too strong and the Balmer jump to be washed out by the stellar Balmer break. This tension makes it impossible to satisfy both the continuum shape from the PRISM and the two-component morphology implied from the photometry with this model.

\subsection{AGN model with stellar emission and DLA}
\label{subsec:agn_star_dla}

\begin{figure*}
	\includegraphics[width=\textwidth]{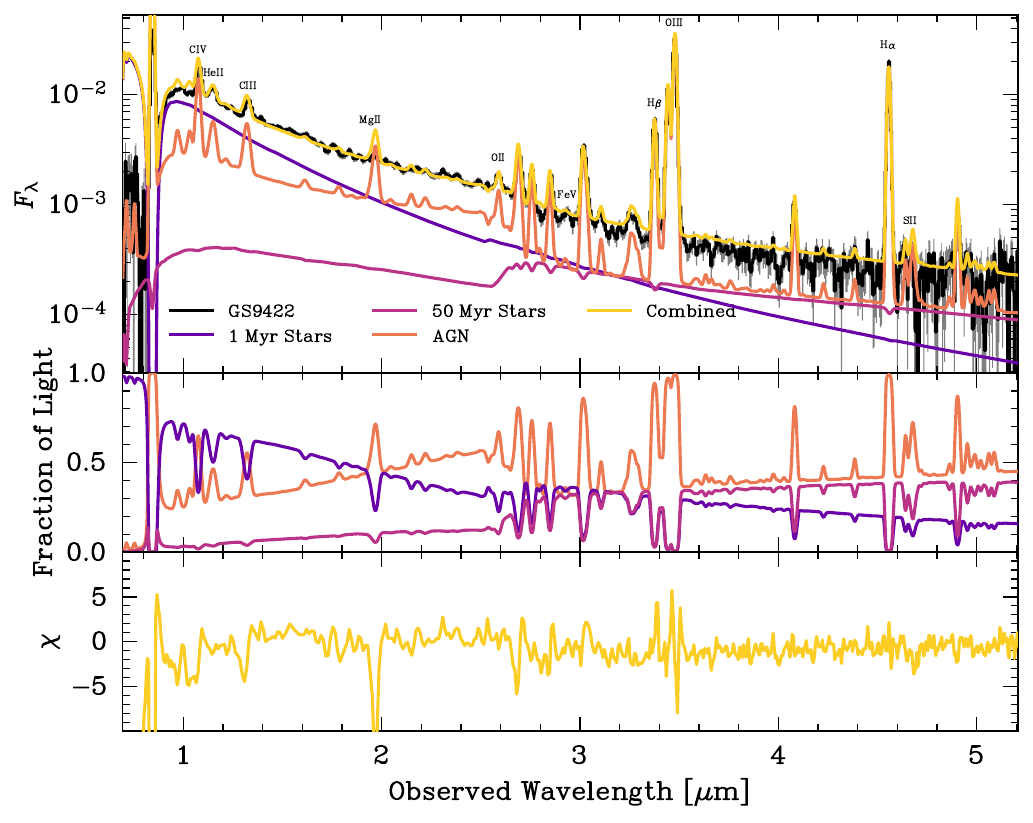}
    \caption{\textit{Top:} The JWST/NIRSpec prism spectrum of GS9422 (black) compared to a three-component AGN-stellar model (yellow). The three-component model is made of the AGN model (orange), a 1 Myr population stellar population (purple), and the maximum contribution from 50 Myr stellar population obscured by dust (dark magenta). The 1 Myr population is used here to show how introducing a steep slope can increase the freedom to add optical flux from older stars while still maintaining the overall continuum shape. However, the significant UV flux of these stars outshines the nebular two-photon turnover, requiring a DLA to recover the observed UV continuum shape. \textit{Middle:} fractional contribution to the emission from the AGN (orange), young stars (purple) and older stars (dark magenta). \textit{Bottom:} absolute value of $\chi$ between data and model. We find a median $|\chi|$ of 1.00, which indicates that the model is able to reproduce the data well. }
    \label{fig:third_sed_model}
\end{figure*}

\begin{figure*}
	\includegraphics[width=\textwidth]{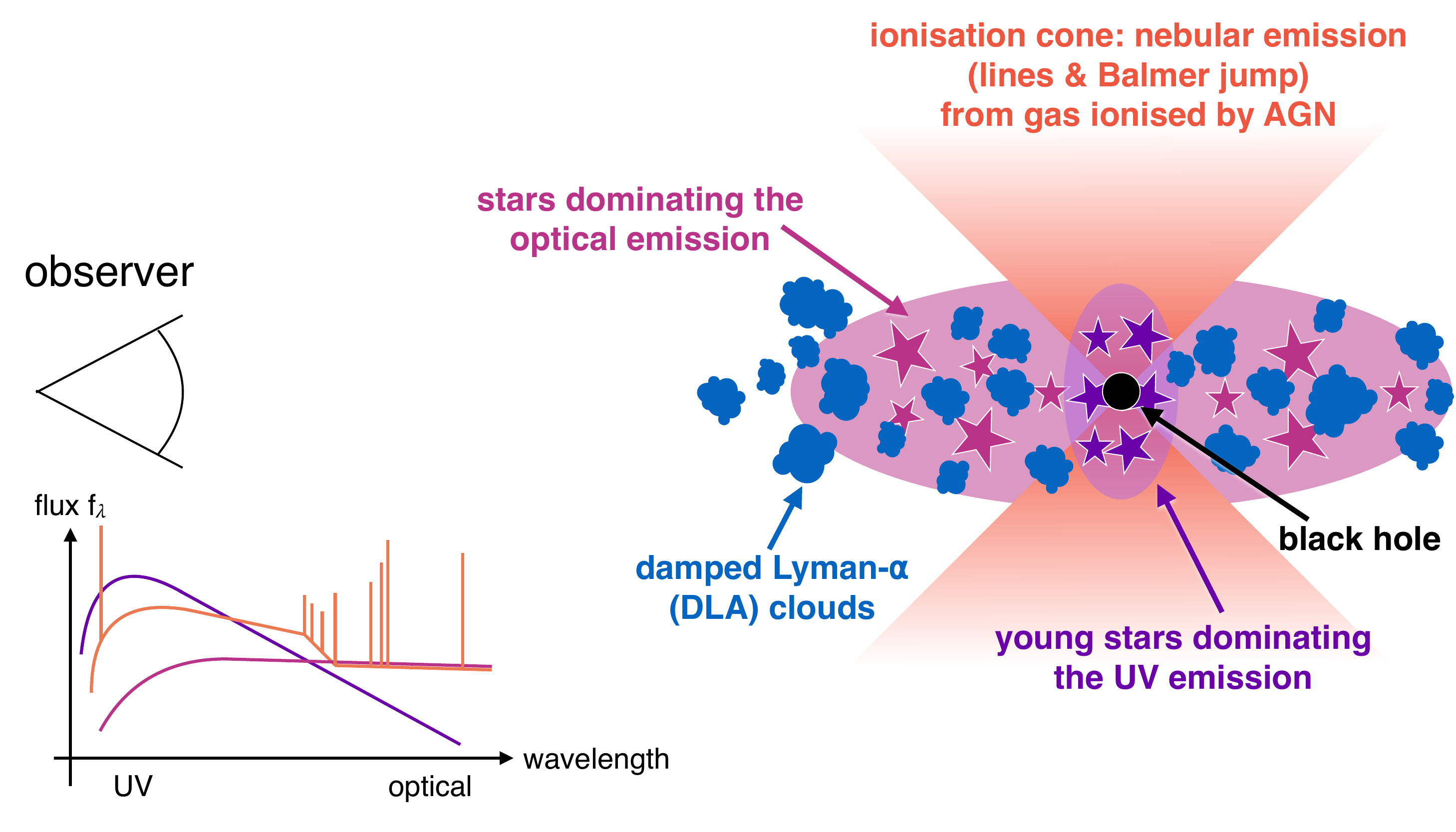}
    \caption{Global schematic of the AGN ionisation cone in GS9422, with a compact young stellar component and an extended older component (not to scale). The AGN ionisation cone gives rise to the Balmer jump and the observed emission lines. The UV is dominated by the young stellar component, where the UV turn-over is caused by DLA clouds, which could simply be an extended \HI disc that we are looking through given our edge-on view of the galaxy, and/or cold, dense gas streams piercing the halo into the central stellar disc. The disc, dominated by older stars, contributes significantly to the rest-frame optical continuum emission. This setup self-consistently explains the spectrum (Balmer jump, strength of emission lines, and emission line ratios) and the morphological variation as a function of wavelength (in particular morphological differences between emission lines, rest-UV continuum and rest-optical continuum emission). }
    \label{fig:cartoon}
\end{figure*}

As highlighted above, the pure AGN model (Section~\ref{subsec:pure_agn}) fits the spectrum overall well, but two issues remain. Firstly, the wavelength dependence of the structure requires that the emission lines and the rest-optical continuum are emitted by different regions. Secondly, the \OII\ doublet is poorly fitted by the pure AGN model. We tried to address the wavelength-dependent morphological demands in Section~\ref{subsec:agn_star}, where we added an old stellar population which is responsible for the offset optical continuum emission. However, this model can only sustain $\sim20$\% of the optical flux from old stars, which is insufficient, and the issue of the poorly fit \OII\ doublet remained.

We can address both issues by adding a young stellar population. In this scenario, the turnover in the UV at $\lambda_{\rm rest}\approx1430$ \AA\ is not caused anymore by the two-photon processes, but by a local DLA, which is associated with the ISM of the galaxy and/or the outer gaseous disc of the galaxy. Since the young stellar component outshines the AGN in the rest-UV, this leads to more flexibility in the modelling of the AGN. Specifically, we can now reduce the two-photon continuum of the AGN (by using a different stopping condition which terminates the calculation earlier) without sacrificing the quality of our fit to the continuum. This leads both to a much better fit to the \OII\ doublet and to additional freedom to increase the strength stellar continuum and, hence, match the morphology.

In particular, we add a young stellar population with a steep UV continuum slope that outshines the aforementioned AGN model, similar to the scenario that can explain the SED of the ``little red dots'' \citep[LRDs;][]{matthee24}. An alternative possibility is to have leaking UV emission from the obscured AGN \citep[e.g.,][]{greene24}, but we do not consider this further here. For our young stellar population, we use 1\,Myr instantaneous starburst. We do not explicitly include the nebular emission from the 1\,Myr stellar population in this particular model. However, since the power-law slope of the AGN ionizing spectrum is a free parameter, the AGN photoionization model can be interpreted as effectively encompassing the joint contribution of both an AGN and a young stellar population. That is, by adjusting the slope appropriately, the combined ionizing spectrum can remain consistent with the observed line ratios. As we have already shown in Fig.~\ref{fig:sed_model_stellar}, the young stars alone are capable of producing similar line diagnostics. While we could redistribute the emission line flux between the AGN and stellar contributions, this would depend on the assumed stopping conditions, especially given that the system likely hosts density-bounded nebulae. In such cases, the relative contributions become underconstrained and somewhat arbitrary. We therefore chose to illustrate one representative and physically plausible configuration, rather than attempt an exact decomposition.

We then change our AGN \texttt{CLOUDY} model stopping condition, which was previously set to ensure the sufficient strength of the nebular continuum. We now set the stopping condition to be based on the \OII/\OIII line ratio. We again note that this leads to density-bounded emission, which is consistent with the non-Case B Balmer line ratios in GS9422 \citep{mcclymont24}. The AGN \texttt{CLOUDY} model is otherwise unchanged.

The two-photon turnover is now outshone by the bright young stars. This means that we require a DLA to create the observed turnover with a damping wing. We find that a neutral hydrogen column density of $\log(N_{\mathrm{HI}}\,\mathrm{cm^{-2})}=22.9\pm0.1$ sufficiently damps the 1\,Myr stellar population in the optical to recreate the observed turnover. While this column density is high, it is consistent with other observed DLAs (see below). The DLA in our scenario could simply be an extended \HI disc that we are looking through given our edge-on view of the galaxy, and/or cold, dense gas streams piercing the halo into the central stellar disc \citep[as in, e.g.,][]{bennett20}.

We find that our modelled stellar populations correspond to masses of $10^{7.5}\mathrm{M}_\odot$ and $10^{8.4}\mathrm{M}_\odot$ for the 1\,Myr and 50\,Myr population respectively. These numbers should not be taken as a fit due to the modelling limitations, however, they do demonstrate that it is possible to accommodate a reasonable stellar mass alongside the AGN-driven nebular emission. We note that, in contrast, the model presented in Section~\ref{subsec:pure_stellar} includes a young stellar component of similar age and mass, but as the sole contributor to the observed emission. Here, instead, the young stellar component is part of a composite emission scenario that may include contributions from an AGN and older stellar populations. In this framework, the physical coordination requirements are relaxed, and the role of the young stars is to augment -- not entirely explain -- the observed line ratios. This makes the model both more flexible and more physically plausible.

The rest-optical continuum in this model approaches the observed flux density. In this case, the old stellar component contributes roughly half of the observed rest-optical light -- the maximum allowed before exceeding the data, but still consistent within uncertainties. Importantly, this model also better reproduces the observed rest-optical morphology: the older stellar population, distributed in a flattened configuration, naturally explains the extended emission seen in the rest-optical imaging. The inclusion of a young stellar component helps preserve the Balmer jump, which would otherwise be diminished by the older stars alone. We also note that the slit-loss correction is optimized for compact sources and may underestimate the true flux density in more extended systems like GS9422, which could contribute to small residual differences. Thus, while this model pushes toward the edge of acceptable parameter space, it remains a physically viable scenario.

We lay out the geometry in Fig.~\ref{fig:cartoon}, which can reproduce both the observed spectrum and the wavelength-dependence of the morphology. In particular, our analysis suggests that GS9422 hosts a young and older stellar population: the centrally concentrated, young stellar population dominates in the UV, while the older stellar population in a disc-like structure dominates in the optical. The DLA absorbs the UV light, giving rise to the UV turnover. The Ly-$\alpha$ emission (and all other emission lines) and the Balmer jump are emitted off-planar and, hence, are unabsorbed by the DLA. This ionisation cone is powered by the AGN, which explains the nebular continuum (e.g., Balmer jump), the emission line strengths and their ratios. The older stellar component in a disc-like configuration explains the observed flattened morphology in the rest-optical bands. 

In the scenario presented here, the nebular continuum — including the Balmer jump — arises from low-density, off-planar ionized gas illuminated by the AGN. Although the AGN is obscured along our direct line of sight, its ionizing radiation can escape along the polar direction and photoionise surrounding gas. This results in extended emission line and continuum features, consistent with an ionization cone. Such off-planar structures are observed in many local AGN host galaxies (e.g., NGC 1068) and provide a natural explanation for how nebular continuum emission can remain visible even in systems classified as Type 2. In GS9422, this off-planar nebular continuum is not blocked by the DLA absorber, allowing it to contribute significantly to the observed rest-optical continuum. While nebular continuum is more commonly studied in Type 1 AGN due to the ease of detection, its physical production is not restricted to such geometries. Given the compactness of the emission region and the observed Balmer jump strength, this component is competitive with — and in our model, partially dominates over — the older stellar continuum in the rest-optical bands. Therefore, it is not surprising that we find a galaxy with such a configuration also at high redshifts, particularly since GS9422 is one out of hundreds observed as part of the JADES survey.

\subsection{Further assessing the DLA scenario}

Some of the scenarios discussed in the previous sections interpret the turnover near Ly$\alpha$ as a local DLA, i.e., associated with the ISM in the outer part of the galactic disc, seen edge-on. Here we discuss the plausibility of the inferred column density.

We first note that several DLAs with column densities approaching $10^{23}~\mathrm{cm}^{-2}$ have been observed at high redshift, specifically with $\log{(N_H/\mathrm{cm}^{-2})}$ up to $23.8 \pm 0.3$ \citep{jakobsson06, prochaska09, watson06, selsing19}, despite their identification being biased towards lower column densities.

We also note that even assuming the most conservative case that the gas is confined to the stellar disc radius, i.e., 180 pc, the observed column would imply a gas density of only $\sim 100~$cm$^{-3}$, which is actually low compared with the gas densities inferred for the compact galaxies found by JWST. This is very conservative, as the bulk of the absorbing gas is likely located in the outer disc, hence at radii larger than the stellar disc. In summary, a scenario in which the ISM of the galaxy seen edge-on, even with modest gas densities, results in column densities even well in excess of $10^{23}~\mathrm{cm}^{-2}$ is quite reasonable.

One can explore what would be the dust extinction associated with the local DLA's gas column density. This can be simply inferred, through the following equation:
\begin{equation}
    \mathrm{A}_{\rm _V,DLA} = \left( \frac{\mathrm{A}_{\rm _V}}{N_{\rm \HI}} \right)_{\rm MW} \cdot \frac{Z_{\rm DLA}}{Z_{\rm MW}}\cdot \frac{\xi_\mathrm{d,DLA}}{\xi_\mathrm{d,MW}}~N_{\rm \HI,DLA},
\end{equation}
where $(\mathrm{A}_{\rm _V}/N_{\HI})_{\rm MW}$ is the extinction-to-gas ratio in the Milky way, i.e. $4.6\times 10^{-22}~\mathrm{cm}^{-2}$ \citep{zhu17}. $Z_{\rm MW}$ is the average metallicity of the ISM in the Milky Way, which is about solar (averaged among most of the lines of sight probed by \citealt{zhu17} and reference therein). $Z_{\rm DLA}$ is the metallicity of the local DLA, which we do not know, but we can conservatively assume to be lower than the metallicity probed by the nebular lines in the ISM of the galaxy. This is likely a conservative assumption, as the bulk of the absorption comes from the outer disc. Therefore, $Z_{\rm DLA} < Z_{\rm ISM} \approx 0.1~Z_\odot$. $\xi_\mathrm{\rm d,MW}=0.46$ is the dust-to-metal mass ratio in the Milky Way \citep{konstantopoulou24}, and $\xi_\mathrm{d,DLA}$ is the dust-to-metal ratio in the DLA that, for high-$z$ absorbers with metallicity similar metallicity to our object, is about is  $0.20^{+0.19}_{-0.13}$ \citep{konstantopoulou24}; the latter is the largest source of uncertainty. We note that since we only have an upper limit on the DLA metallicity, this implies that also $\xi_\mathrm{\rm d,DLA}$ is an upper limit.

Combining the information above, we can, therefore only infer an upper limit on the amount of dust extinction as
\begin{equation}
\mathrm{A}_{\rm _V,DLA} < 1.25^{+1.40}_{-1.02}
\end{equation}
Essentially, the extinction associated with the DLA can well be as low as the dust attenuation inferred from the spectral fitting discussed above ($\mathrm{A}_{\rm _V}=0.1-0.5$). While dust attenuation is necessarily higher than extinction (the latter does not include scattering and re-emission into the line of sight), the uncertainties in the above calculation are much larger than this effect. We further note that the vast majority of the DLAs observed at high redshift with the largest column densities ($N_H>10^{22}~\mathrm{cm}^{-2}$) have $\mathrm{A}_{\rm _V}<0.2$ \citep{heintz24, deugenio24_C}.

\section{Summary \& conclusions}
\label{sec:conclusions}

The nebular continuum and nebular emission lines contain a wealth of information regarding galaxy properties and physical processes taking place within galaxies. Here we focus on a galaxy (GS9422) at the tail end of cosmic reionisation at a redshift $z_{\rm spec}=5.943$. Because of the observed Balmer jump and steep UV continuum turnover, \citet{cameron24} suggest that GS9422 is completely dominated by the nebular continuum, and that the UV turnover is caused by two-photon emission. They conclude that this would require the galaxy to be dominated by low-metallicity stars of $\gtrsim50~M_{\odot}$, which can only be produced by an IMF $10-30\times$ more top-heavy than those typically assumed. In this work, we use detailed imaging and spectroscopic data from the JADES and JEMS surveys for this galaxy, GS9422, in order to resolve the nature of the galaxy and putative nebular emission.

The photometry confirms the Balmer jump detected in the prism spectrum, and we find clear variations in the galaxy morphology as a function of wavelength (Fig.~\ref{fig:sed}). Specifically, we find that the rest-frame optical continuum emission is consistent with an elongated disc, while the rest-frame UV is centrally concentrated. On the other hand, the emission from the nebular lines extends perpendicular to the disc (Figs.~\ref{fig:stack}, \ref{fig:isophotal} and \ref{fig:pysersic}). This leads to variations of the H$\alpha$+\NII EW across the galaxy, with a higher EW along the poles of the galaxy (cone-like structure; Fig.~\ref{fig:ew_map}).

We model the prism spectrum with a range of approaches. First (Fig.~\ref{fig:sed_model_stellar}), we fit the spectrum with a pure stellar emission model, where young stars (with an age of $1-2$ Myr) drive the nebular continuum emission. A DLA is needed to reproduce the turn-over in the rest-UV. While this model is able to reproduce the spectrum well without invoking a top-heavy IMF and exotic stars, it is inconsistent with the morphological constraints: it cannot explain the strong morphological difference between nebular emission lines and nebular continuum. 

Motivated by rest-frame UV and optical emission line diagnostics that are fully consistent with AGN (Figs.~\ref{fig:eml_ratios} and \ref{fig:UV_diagnostics}), we explore models with AGN. Similar to the pure stellar model, the pure type-2 AGN model is able to reproduce the spectrum well (Fig.~\ref{fig:sed_model}). This time, no DLA is needed, and the UV turnover can be modelled by the two-photon process. This model, however, does not account for the morphological differences in the rest-UV, rest-optical and emission lines. Adding an older stellar component to this model does not solve this problem, since the continuum is dominated by the nebular emission of the AGN (Fig.~\ref{fig:combo_sed_model}). 

Therefore, our preferred model includes a galaxy with an obscured, type-2 AGN and DLA clouds associated with the galaxy's ISM or its outskirts (Figs.~\ref{fig:third_sed_model} and \ref{fig:cartoon}). This model can explain both the spectrum and the wavelength dependence of the morphology, without resorting to IMF changes. The AGN powers the off-planar nebular emission, giving rise to the Balmer jump and the emission lines, while a centrally concentrated, young stellar component dominates the rest-UV emission and together with the DLA clouds leads to a turn-over of the spectrum towards shorter wavelengths. The older stellar component in a disc-like configuration explains the observed flattened morphology in the rest-optical. In summary, we find that GS9422 is consistent with being a normal galaxy with an AGN -- a simple scenario that can explain its properties, without the need for exotic or extraordinary stellar populations. 

This work highlights the importance of combining imaging (JWST/NIRCam) and spectroscopic data (JWST/NIRSpec), and the great potential for NIRSpec/IFU observations. While spectroscopy delivers detailed insights into a range of galaxy properties, spatially resolved information helps with breaking some of the degeneracies regarding the mechanisms that can power the observed emission.

\section*{Acknowledgements}

We would like to thank Alex Cameron, Harley Katz, Aayush Saxena and Andy Bunker for the clarifications about their work and constructive feedback. We are grateful to Annalisa De Cia for useful suggestions.

The authors acknowledge use of the lux supercomputer at UC Santa Cruz, funded by NSF MRI grant AST 1828315. ST acknowledges support by the Royal Society Research Grant G125142. WM acknowledges the Technology Facilities Council (STFC) Center for Doctoral Training (CDT) in Data intensive science at the University of Cambridge (STFC grant number 2742605) for a PhD studentship. JS, RM, XJ, FD, JW CSS,  acknowledges support by the Science and Technology Facilities Council (STFC), by the ERC through Advanced Grant 695671 “QUENCH”, and by the UKRI Frontier Research grant RISEandFALL. RM also acknowledges funding from a research professorship from the Royal Society. NCV acknowledges support from the Charles and Julia Henry Fund through the Henry Fellowship. The research of CCW is supported by NOIRLab, which is managed by the Association of Universities for Research in Astronomy (AURA) under a cooperative agreement with the National Science Foundation. JC acknowledges funding from the "FirstGalaxies" Advanced Grant from the European Research Council (ERC) under the European Union’s Horizon 2020 research and innovation programme (Grant agreement No. 789056). SC acknowledges support by European Union’s HE ERC Starting Grant No. 101040227 - WINGS. DP acknowledges support by the Huo Family Foundation through a P.C. Ho PhD Studentship. BER, BJ and CNAW acknowledge support from the NIRCam Science Team contract to the University of Arizona, NAS5-02015, and JWST Program 3215. MSS acknowledges support by the Science and Technology Facilities Council (STFC) grant ST/V506709/1. H{\"U} gratefully acknowledges support by the Isaac Newton Trust and by the Kavli Foundation through a Newton-Kavli Junior Fellowship.

%%%%%%%%%%%%%%%%%%%%%%%%%%%%%%%%%%%%%%%%%%%%%%%%%%
\section*{Data Availability}

Fully reduced NIRCam images and NIRSpec spectra are publicly available on MAST (\url{https://archive.stsci.edu/hlsp/jades}), with \doi{10.17909/8tdj-8n28}, \doi{10.17909/z2gw-mk31}, and \doi{10.17909/fsc4-dt61} \citep{rieke23, bunker24, eisenstein23_jades, williams23_jems}.



\bibliographystyle{mnras}
\bibliography{library} % if your bibtex file is called example.bib

%\begin{thebibliography}{99}
%\end{thebibliography}

%%%%%%%%%%%%%%%%%%%%%%%%%%%%%%%%%%%%%%%%%%%%%%%%%%

%%%%%%%%%%%%%%%%% APPENDICES %%%%%%%%%%%%%%%%%%%%%

\appendix

\section{\HeII$\lambda$1640 line}
\label{app:HeII}

\begin{figure}
	\includegraphics[width=\linewidth]{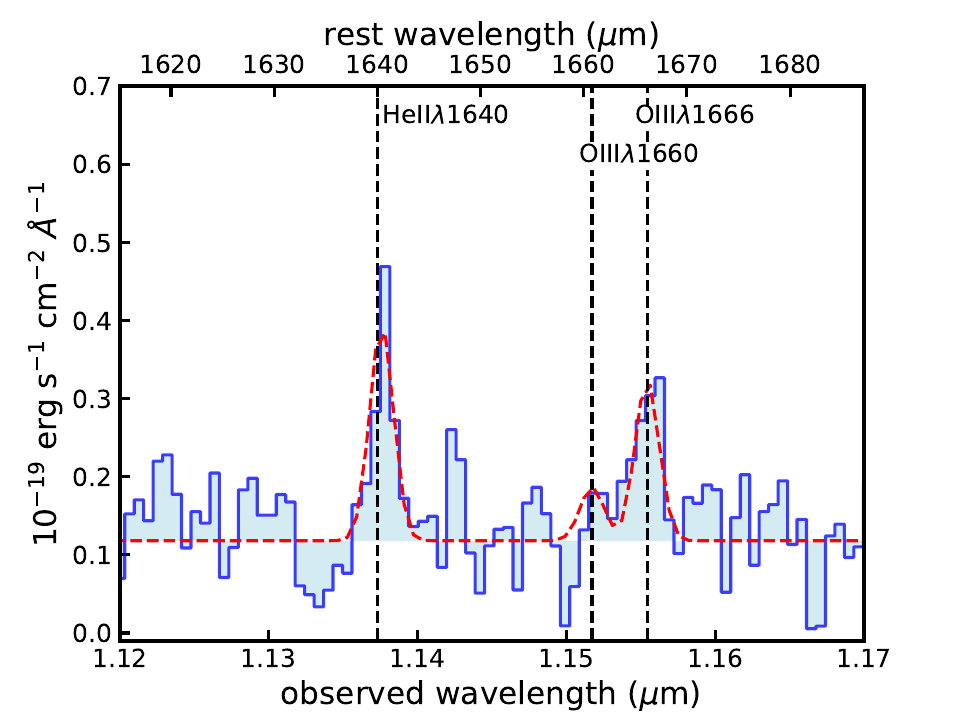}
    \caption{Spectrum of the HeII$\lambda$1640 and OIII]$\lambda\lambda$1660,66 in the G140M grating. The plot shows the data (blue line) and the best fit (red line) to the observed spectrum.}
    \label{fig_app:HeII}
\end{figure}

We present the spectrum of the \HeII$\lambda$1640 and OIII]$\lambda\lambda$1660,66 in the G140M grating in Fig.~\ref{fig_app:HeII}. The plot shows the data and the best fit to the observed spectrum. We find a flux for \HeII$\lambda$1640 of $(5.5\pm0.6)\times10^{-19}~{\rm erg/s/cm^{2}}$, while \citet{cameron24} found $(5.1\pm0.9)\times10^{-19}~{\rm erg/s/cm^{2}}$, which is consistent within the uncertainty.

\section*{Affiliations}
\noindent
{\it
$^{1}$Kavli Institute for Cosmology, University of Cambridge, Madingley Road, Cambridge, CB3 0HA, UK\\
$^{2}$Cavendish Laboratory, University of Cambridge, 19 JJ Thomson Avenue, Cambridge, CB3 0HE, UK\\
$^{3}$Sorbonne Universit\'e, CNRS, UMR 7095, Institut d'Astrophysique de Paris, 98 bis bd Arago, 75014 Paris, France\\
$^{4}$Steward Observatory, University of Arizona, 933 N. Cherry Avenue, Tucson, AZ 85721, USA\\
$^{5}$NSF’s National Optical-Infrared Astronomy Research Laboratory, 950 North Cherry Avenue, Tucson, AZ 85719, USA\\
$^{6}$European Space Agency (ESA), European Space Astronomy Centre (ESAC), Camino Bajo del Castillo s/n, 28692 Villanueva de la Ca\~nada, Madrid, Spain\\
$^{7}$Scuola Normale Superiore, Piazza dei Cavalieri 7, I-56126 Pisa, Italy\\
$^{8}$Department of Physics, University of Oxford, Denys Wilkinson Building, Keble Road, Oxford OX1 3RH, UK\\
$^{9}$European Southern Observatory, Karl-Schwarzschild-Strasse 2, 85748 Garching, Germany\\
$^{10}$Center for Astrophysics $|$ Harvard \& Smithsonian, 60 Garden St., Cambridge MA 02138 USA\\
$^{11}$Department of Astronomy \& Astrophysics, The Pennsylvania State University, University Park, PA 16802, USA\\
$^{12}$Institute for Computational \& Data Sciences, The Pennsylvania State University, University Park, PA 16802, USA\\
$^{13}$Institute for Gravitation and the Cosmos, The Pennsylvania State University, University Park, PA 16802, USA\\
$^{14}$Department of Astronomy, University of Wisconsin-Madison, 475 N. Charter St., Madison, WI 53706 USA\\
$^{15}$Department of Astronomy and Astrophysics University of California, Santa Cruz, 1156 High Street, Santa Cruz CA 96054, USA\\
$^{16}$Department of Astronomy and Astrophysics, University of California, Santa Cruz, 1156 High Street, Santa Cruz, CA 95064, USA\\
$^{17}$Centro de Astrobiología (CAB), CSIC-INTA, Ctra. de Ajalvir km 4, Torrej\'on de Ardoz, E-28850, Madrid, Spain\\
$^{18}$Centre for Astrophysics Research, Department of Physics, Astronomy and Mathematics, University of Hertfordshire, Hatfield AL10 9AB, UK\\
$^{19}$NRC Herzberg, 5071 West Saanich Rd, Victoria, BC V9E 2E7, Canada
}

%%%%%%%%%%%%%%%%%%%%%%%%%%%%%%%%%%%%%%%%%%%%%%%%%%

% Don't change these lines
\bsp	% typesetting comment
\label{lastpage}
\end{document}